\documentclass[lettersize,journal]{IEEEtran}
\usepackage{amsmath,amsfonts}
\usepackage{algpseudocode}
\usepackage{threeparttable}
\usepackage{amsmath}
\usepackage{amssymb}
\usepackage{mathtools}
\usepackage{algorithm}
\usepackage{array}
\usepackage[caption=false,font=normalsize,labelfont=sf,textfont=sf]{subfig}
\usepackage{textcomp}
\usepackage{stfloats}
\usepackage{soul}
\usepackage{enumitem}
\usepackage{accessibility}
\usepackage{url}
\usepackage{verbatim}
\usepackage{graphicx}
\usepackage{xurl}
\DeclareUrlCommand{\code}{\urlstyle{tt}}
\usepackage{cite}
\usepackage{booktabs}
\usepackage{multirow}
\usepackage{scalerel}
\usepackage{tikz}

\usepackage{hyperref}

\newcommand{\codde}[1]{\nolinkurl{#1}}
\usetikzlibrary{svg.path}

\definecolor{orcidlogocol}{HTML}{A6CE39}
\tikzset{
  orcidlogo/.pic={
    \fill[orcidlogocol] svg{M256,128c0,70.7-57.3,128-128,128C57.3,256,0,198.7,0,128C0,57.3,57.3,0,128,0C198.7,0,256,57.3,256,128z};
    \fill[white] svg{M86.3,186.2H70.9V79.1h15.4v48.4V186.2z}
                 svg{M108.9,79.1h41.6c39.6,0,57,28.3,57,53.6c0,27.5-21.5,53.6-56.8,53.6h-41.8V79.1z M124.3,172.4h24.5c34.9,0,42.9-26.5,42.9-39.7c0-21.5-13.7-39.7-43.7-39.7h-23.7V172.4z}
                 svg{M88.7,56.8c0,5.5-4.5,10.1-10.1,10.1c-5.6,0-10.1-4.6-10.1-10.1c0-5.6,4.5-10.1,10.1-10.1C84.2,46.7,88.7,51.3,88.7,56.8z};
  }
}

\newcommand\orcidicon[1]{\href{https://orcid.org/#1}{\mbox{\scalerel*{
\begin{tikzpicture}[yscale=-1,transform shape]
\pic{orcidlogo};
\end{tikzpicture}
}{|}}}}
\usepackage{hyperref}
\begin{document}

\title{Sim-FA: A GPGPU Simulator Framework for Fine-Grained Asynchronous Pipeline Analysis}

\author{Zhongchun~Zhou\orcidicon{0009-0000-7037-7418},~Yuhang~Gu\orcidicon{0009-0001-5040-9424},~Chengtao~Lai\orcidicon{0000-0002-9547-9653},~Ya~Wang\orcidicon{0009-0009-1114-7159},~Zeyu~Han\orcidicon{0009-0002-1627-4439},~Wei~Zhang\orcidicon{0000-0002-7622-6714},~\IEEEmembership{Fellow,~IEEE},~and~Jun~Liu\orcidicon{0009-0008-3046-1779}%

\thanks{Z. Zhou, C. Lai, and Y. Wang are with the Department of Electronic and Computer Engineering, 
The Hong Kong University of Science and Technology, Clear Water Bay, Kowloon, Hong Kong 
(e-mail: zzhouch@connect.ust.hk; claiaf@connect.ust.hk; ywangmu@connect.ust.hk).}%
\thanks{Y. Gu is with the School of Electronic Science and Engineering, Southeast University, Nanjing, Jiangsu, China}
\thanks{Z. Han is with school of Integrated Circuits, Nanjing
University, Nanjing, China}
\thanks{W. Zhang (corresponding author) is with the Department of Electronic and Computer Engineering, 
The Hong Kong University of Science and Technology, Clear Water Bay, Kowloon, Hong Kong 
(e-mail: eeweiz@ust.hk).}%
\thanks{J. Liu is with ZTE corporation, Shenzhen, China}
}



\maketitle

\begin{abstract}
To efficiently support Large Language Models (LLMs), modern GPGPU architectures have introduced new features and programming paradigms, such as warp specialization. These features enable temporal overlap between the producer and consumer, as well as between matrix multiplication and activation function operations, substantially improving performance. To conduct effective AI infrastructure and computer architecture research, cycle-level simulators that support these new features, together with analytical models that faithfully capture workload characteristics, are essential.


In this paper, we build Sim-FA, a WarpGroup-oriented cycle-level simulation framework for Hopper TMA/WGMMA pipelines, achieving a 7.4$\times$ geometric-mean simulation speedup over Accel-Sim. We first develop an operator-independent trace frontend that instruments kernels at the Triton TTGIR level and validates it on 23 GEMM shapes, achieving 5.49\% MAPE against H800, confirming that the simulator core is not tied to any single operator. Because FlashAttention-3 introduces additional complexity beyond standard TMA/WGMMA kernels (asymmetric producer-consumer pipelines, softmax, ping-pong synchronization), we further build an FA3-specialized frontend that achieves 5.7\% MAPE against H800 and 3.28\% MAPE against H20. Within the same framework, SimFA-python serves as an analytical fast path for large-scale design-space exploration where cycle-level simulation is prohibitively slow; validated against cuTile kernels on Blackwell (GB10), it explains why existing analytical models can produce inaccurate traffic estimates.
\end{abstract}

\begin{IEEEkeywords}
GPU, Simulation, Accelerator, Large Language Model.
\end{IEEEkeywords}

\section{Introduction}
Recent industrial experience in LLM deployment has shown the significance of hardware-model co-design~\cite{Deepseek}:
not only should the model architecture fit the hardware, operators must also be deeply optimized to exploit new hardware features. FlashAttention-3~\cite{fa3} is one such critical operator: highly optimized for NVIDIA Hopper GPUs, it fully utilizes TMA (Tensor Memory Accelerator) and WGMMA (WarpGroup MMA)~\cite{luo2025hopper}, together with warp specialization~\cite{cudadma}.

However, this approach has seen slower adoption in academic research.
State-of-the-art features like TMA have long lacked timely support from mainstream simulators such as Accel-Sim~\cite{AccelSim};
its recent 2.0 release~\cite{accelsim2} closes this feature gap but retains a general-purpose instruction-level execution model, which incurs high simulation costs for TMA/WGMMA-intensive ML kernels.
On the other hand, Python-based roofline and analytical models common in academia~\cite{genz, neusight, flat} will face difficulty in analyzing pipeline bubbles, resource contention, and the effectiveness of architectural innovation.
Yet building cycle-accurate LLM simulators faces several challenges. First, the workload size of LLMs make detailed architecture simulation extremely slow. Second, NVIDIA's GPU designs are not open to academia, though recent micro-benchmarking studies have begun to characterize them~\cite{luo2025hopper, blackwell}. Finally, asynchronous features including TMA and WGMMA require fine-grained synchronization primitives and complex dependency chains.
The key challenge is therefore not merely to support Hopper asynchronous primitives, but to identify which microarchitectural state must be retained to reproduce their performance-critical timing interactions at substantially lower simulation cost. \textbf{Sim-FA addresses this challenge by treating WarpGroups, rather than individual instructions, as the primary logical execution unit for asynchronous ML kernels.}

Additionally, we utilize the micro-benchmark results to configure our simulator and omit unnecessary detailed pipeline modeling that is irrelevant to the problem, thereby accelerating simulation. This work makes the following contributions:
\begin{itemize}
\item To our knowledge, Sim-FA is the first WarpGroup-oriented cycle-level simulation abstraction for Hopper asynchronous ML kernels.\footnote{An early version of this work was publicly released on arXiv on May 1, 2026 (arXiv:2605.00555).}  Rather than reproducing all low-level pipeline state of a general-purpose GPU simulator, Sim-FA explicitly preserves the timing-visible interactions among TMA transfers, WGMMA execution, barrier synchronization, producer-consumer scheduling, and the memory hierarchy. Sim-FA achieves 5.7\% MAPE on FlashAttention-3. In a direct simulation-cost comparison on 23 operator-equivalent GEMM workloads, it is faster than Accel-Sim 2.0 in 22 of 23 cases, with a 7.4$\times$ geometric-mean simulation speedup and 5--23$\times$ lower host-memory footprint.
\item We build an operator-agnostic trace frontend that instruments kernels at the TTGIR (Triton GPU IR) level, requiring no operator-specific instrumentation or event decoding. Validated on 23 GEMM shapes at 5.49\% MAPE with the same simulator configuration used for FA3, it demonstrates that the WarpGroup-level core model generalizes beyond attention.
\item A detailed mathematical analysis of system traffic for FlashAttention is presented, and the actual traffic statistics profiled from the NVIDIA Nsight Compute software are used for comparison. It reveals the problems with current analytical tools.
\end{itemize}

\section{Background and Motivation}
\label{background}
This section introduces GPU asynchrony, illustrates its use through FlashAttention, and motivates Sim-FA by discussing the limitations of existing analytical and cycle-level performance models.


\subsection{Features to Support Asynchrony in GPUs}
\label{advanced_feature}
\subsubsection{TMA}
Proposed in the Hopper architecture, its efficacy is twofold. (1) It accepts a TensorMap object that stores the base pointer, data type, shape (1D to 5D), and strides of a tensor. At runtime, TMA uses this metadata to generate addresses in hardware and fetch bulk data from global memory to shared memory. This frees the CUDA cores from the index calculation of every element and saves registers. (2) TMA works asynchronously with the computing units, enabling explicit memory latency hiding.

\subsubsection{mbarrier} It is a hardware-accelerated synchronization object stored in shared memory. It features a split arrive/wait barrier, which can be used to implement fine-grained thread control, a producer-consumer pipeline, and divergent code patterns in CUDA. It is intended to be paired with asynchronous memory access instructions such as TMA.

\subsubsection{Warp Specialization}
\label{WarpSpecialization}
Unlike conventional SIMT execution that relies on warp-level context switching to hide memory latency, warp specialization partitions warps within a thread block to perform distinct tasks (e.g., data fetching versus computation), enabling explicit producer-consumer pipelining.
Each pipeline role is statically assigned to one or more WarpGroups (4 aligned warps); FlashAttention-3, for instance, uses 1 producer and 2 consumer WarpGroups. WGMMA likewise relies on asynchronous primitives (\texttt{wgmma.commit\_group}/\texttt{wait\_group}), and a set of further synchronization primitives (Table~\ref{tab:simfa-isa}) coordinates execution status across WarpGroups.


\subsection{FlashAttention and Asynchrony}
As the row-wise softmax in multi-head attention hinders fusion, $QK^T$ was conventionally materialized in global memory. FlashAttention leverages online softmax to store and update the attention score tiles on-chip, avoiding expensive memory round-trips. 

As the Hopper architecture systematically introduces asynchrony through TMA and WGMMA, FlashAttention-3 (FA3) defines a warp-specialized software pipelining scheme that exploits asynchrony by splitting producers and consumers into separate warps. As softmax mainly uses CUDA cores and MUFU (Multi-function Units), it can also be overlapped with Tensor Core execution. In addition to such inter-WarpGroup pipelining, it also proposes an intra-WarpGroup overlapping scheme that further hides softmax execution time by introducing additional pipeline buffers in the register file. However, this demand for registers conflicts with the use of larger block sizes, another optimization idea. The final pipeline stages and block sizes are determined through profiling.

FlashAttention-4 inherits this design on Blackwell. As asynchrony is not yet well supported by existing simulators, this work targets Hopper and FlashAttention-3; GEMM and other workloads use the same primitives, as Section~\ref{generic_frontend} validates on 23 shapes.




\subsection{Motivations}
Our work is motivated by the weaknesses and limitations in current analytical models and cycle-level simulators for GPUs:
\begin{itemize}[leftmargin=0.6cm]
    \item Performance-aligned analytical models can be deceptive.
    As we will show later, the final result can still appear accurate when the number of DRAM accesses is severely underestimated.
    Despite their close approximation near well-balanced real-world design points, in design space exploration problems (a typical scenario for such models) where many candidate design points are prevalent, the biases can become apparent.
    \item Studies involving cycle-level simulators also face their own limitations. Some employ unoptimized programs as their benchmarks \cite{LCM} (which is also common in analytical model studies), limiting their potential impact. In terms of architectural modeling, although the recent Accel-Sim 2.0 release \cite{accelsim2} adds Hopper TMA/WGMMA support, its general-purpose instruction-level abstraction retains substantial per-instruction simulation state and execution cost for TMA/WGMMA-intensive kernels, as quantified in Section~\ref{sim_characteristic}.
\end{itemize}

Our work therefore \textbf{places more emphasis on cycle-level simulation}, especially on Hopper asynchrony, while using a corrected analytical model as a substitute for cases whose simulation time would be prohibitive.


\section{Theoretical Framework for FlashAttention Performance Analysis}
\subsection{Problem Formulation and Notation}
In this section, we will derive a theoretical performance model for FlashAttention. Existing analytical models often rely on ideal traffic estimation (Section~\ref{ideal}), significantly underestimating actual DRAM traffic under hardware constraints (Section~\ref{real}). \textbf{The oversimplified calculation in these models can mislead future chip design decisions.} Below we calculate the compute complexity, L2 cache traffic, and DRAM traffic under various situations, thereby offering a theoretical reference for understanding actual hardware designs and trade-offs. The results are validated against actual NVIDIA GPU profiling results in Section \ref{analytical_exp}. 
The analytical model SimFA-python is the fast path of the same framework: it shares the traffic formulation of the cycle-level core but trades fidelity for speed, enabling design-space exploration (DSE).
The related notations and parameters are defined in Table~\ref{tab:notations}. For simplicity, the formulas in this section are for non-causal attention mechanisms.
\begin{table}[t]
  \caption{Summary of Notations and Parameters}
  \label{tab:notations}
  \vspace*{-1\baselineskip}
  \begin{tabular}{l l}
    \toprule
    \textbf{Symbol} & \textbf{Description} \\
    \midrule
    $B$ & Batch size \\
    $L, S$ & Sequence lengths of Query and KV, respectively \\
    $H_{KV}$ & Number of KV heads \\
    $G$ & Query Group Size (Number of Q heads per KV head) \\
    $D$ & Head dimension \\
    $T_M$ & Tile size along the M (Query) dimension \\
    $P$ & Precision size in bytes (e.g., 2 for FP16) \\
    $N_{\mathrm{SM}}$ & Number of Streaming Multiprocessors (SMs) on GPU \\
    $O_{\mathrm{limit}}$ & Max concurrent thread blocks per SM (Occupancy limit) \\
    \bottomrule
  \end{tabular}
\end{table}
\subsection{Computational Complexity (FLOPs)}
FlashAttention comprises two matrix multiplication operations, $\mathbf{Q} \mathbf{K}^T$, and $\text{softmax}(\mathbf{Q} \mathbf{K}^T) \cdot \mathbf{V}$, along with a softmax operator. For each Q-head, the shape of $\mathbf{Q} \mathbf{K}^T$ is $(L \times D) \times (D \times S) \xrightarrow{} (L \times S) $, and the shape of $\text{softmax}(\mathbf{Q} \mathbf{K}^T) \cdot \mathbf{V}$ is $(L \times S) \times (S \times D) \xrightarrow{} (L \times D) $.
Therefore, the FLOPs for each Q-head are $2  (L \cdot D \cdot S) + 2  (L \cdot S \cdot D)$. 

Since there are $B \cdot (H_{KV} \cdot G)$ Q-heads, the total FLOPs for the non-causal attention mechanism can be formulated as:
\begin{equation}
    \mathcal{F}_{\mathrm{total}} = 4 B \cdot (H_{KV} \cdot G) \cdot L \cdot S \cdot D
    \label{eq:flops}
\end{equation}
\subsection{L2 Cache Access Demand}
We analyze L2 cache traffic at the thread block level, where each block is assigned a single $T_M \times D$ Q-tile / O-tile. Under this model, a thread block loads its Q-tile and writes the resulting O-tile once. To complete the computation, it must traverse the entire K/V head, each with $S \times D$ elements.


Based on this tiling strategy, the number of thread blocks equals $N_{blocks} = B \cdot (G\cdot H_{KV}) \cdot \lceil \frac{L}{T_M} \rceil$. We calculate the total memory traffic issued by all thread blocks, which corresponds to the traffic sent to the L2 cache. For all thread blocks, the traffic contributed by Q-tile reads and O-tile writes equals $2N_{blocks} \cdot T_M \cdot D \approx 2B \cdot (G\cdot H_{KV}) \cdot L \cdot D$, while that of K-head and V-head equals $2N_{blocks} \cdot S \cdot D$.
The total L2 traffic is:
\begin{equation}
    \mathcal{M}_{L2} = P \cdot B \cdot (H_{KV} \cdot G) \cdot D \cdot \left( 2L + \left\lceil \frac{L}{T_M} \right\rceil \cdot 2S \right) \quad (\text{Bytes})
    \label{eq:l2}
\end{equation}
\subsection{Theoretical Minimum DRAM Traffic (Ideal Cache)}
\label{ideal}
If the L2 cache is ideal and large enough, all duplicated accesses to K and V will be merged in the L2 cache (cache hit or MSHR hit). From the DRAM's perspective, the K and V tensors will be loaded only once. The main data traffic can be categorized into four types: (1) Reading the complete Q tensor; (2) Reading the complete K tensor; (3) Reading the complete V tensor; (4) Writing the complete O tensor. 
Therefore, the total DRAM traffic is:
\begin{equation}
    \mathcal{M}_{\text{DRAM}}^{\text{ideal}} = P \cdot B \cdot D \cdot \left( 2 \cdot (H_{KV} \cdot G) \cdot L + 2 \cdot H_{KV} \cdot S \right)
    \label{eq:ideal}
\end{equation}

Ideal caching requires the effective L2 capacity to accommodate a single K head and a single V head. We can process a single KV-head at a time using L2 cache swizzling. Consequently, we can formulate the condition below:
\begin{equation}
    \text{Size}_{\text{L2}} > 2P \cdot S \cdot D
    \label{eq:cond}
\end{equation}
\subsection{Realistic DRAM Traffic}
\label{real}
When the condition (\ref{eq:cond}) cannot be satisfied, Eq. (\ref{eq:ideal}) can no longer be used to predict the DRAM traffic. We model the computing process as a sequence of waves. The thread blocks within the same KV head are dispatched to fully occupy the GPU in one or more waves. We can classify the reuse of K and V tensors into two categories: (1) Reuse across SMs/thread blocks within the same wave. (2) Reuse across different waves.

Under realistic hardware constraints, limited effective L2 capacity prevents all reusable K/V data from remaining resident in the cache. We model this using a concurrency-aware wave model. The number of memory waves (passes) required to process one KV group is:
\begin{equation}
    \mathcal{W}_{\mathrm{grp}} = \max\left( 1, \left\lceil \frac{G \cdot \lceil L/T_M \rceil}{N_{\mathrm{SM}} \cdot O_{\mathrm{limit}}} \right\rceil \right)
\end{equation}

Consequently, the estimated realistic DRAM traffic is:
\begin{equation}
\begin{split}
    \mathcal{M}_{\mathrm{DRAM}}^{\mathrm{real}} =\; & \underbrace{2 \cdot P \cdot B \cdot (H_{KV} \cdot G) \cdot L \cdot D}_{\text{Traffic of Q and O (Base)}} \\
    & + \underbrace{\left( 2 \cdot P \cdot B \cdot H_{KV} \cdot S \cdot D \right)}_{\text{Size of KV}} \times \mathcal{W}_{\mathrm{grp}}
\end{split}
\label{eq:dram_real}
\end{equation}

\subsection{Memory Traffic Ratio Analysis}

To evaluate the effectiveness of the L2 cache in filtering memory requests, we define the L2-to-DRAM Traffic Ratio ($\mathcal{R}_{\text{traffic}}$). This metric reflects the data reuse rate achieved by the memory hierarchy:
\begin{equation}
    \mathcal{R}_{\text{traffic}} = \frac{\mathcal{M}_{L2}}{\mathcal{M}_{\text{DRAM}}^{\text{real}}}
    \label{eq:r_traffic}
\end{equation}
A higher \(R_{\mathrm{traffic}}\) indicates more requests are filtered by the L2, and vice versa.

According to Eq. ~\eqref{eq:l2}, when the sequence length is long, the traffic from K and V is dominant since K/V traffic is $O(L \cdot S)$ compared to the traffic from Q and O, which is $O(L)$.
From Eq. ~\eqref{eq:l2}:
\begin{equation}
    \mathcal{M}_{L2} \approx 2 \cdot P \cdot B \cdot (H_{KV} \cdot G) \cdot D \cdot \frac{L}{T_M} \cdot S \quad (\text{Bytes})
    \label{eq:l2_est}
\end{equation}
Similarly,
\begin{equation}
\begin{split}
    \mathcal{M}_{\mathrm{DRAM}}^{\mathrm{real}} \approx \left( 2 \cdot P \cdot B \cdot H_{KV} \cdot S \cdot D \right) \cdot \left( \frac{G \cdot L}{N_{SM}\cdot T_M \cdot O_{limit}} \right)
\end{split}
\label{eq:dram_real_est}
\end{equation}
From Eq.~\eqref{eq:r_traffic}, \eqref{eq:l2_est}, and~\eqref{eq:dram_real_est}:
\begin{equation}
    \mathcal{R}_{\text{traffic}} \approx N_{SM} \cdot O_{limit}
    \label{eq:r_traffic_est}
\end{equation}
Server-class GPUs typically feature tens of streaming multiprocessors, but the bandwidth ratio between the L2 cache and DRAM is usually less than 10. Therefore, the L2 cache is more likely to become the bottleneck than DRAM bandwidth.


\section{Sim-FA: Simulator Core and Memory Hierarchy}
\label{core}
\label{async_pipe}

\begin{figure*}[!t]
\centerline{\includegraphics[width=1\linewidth]{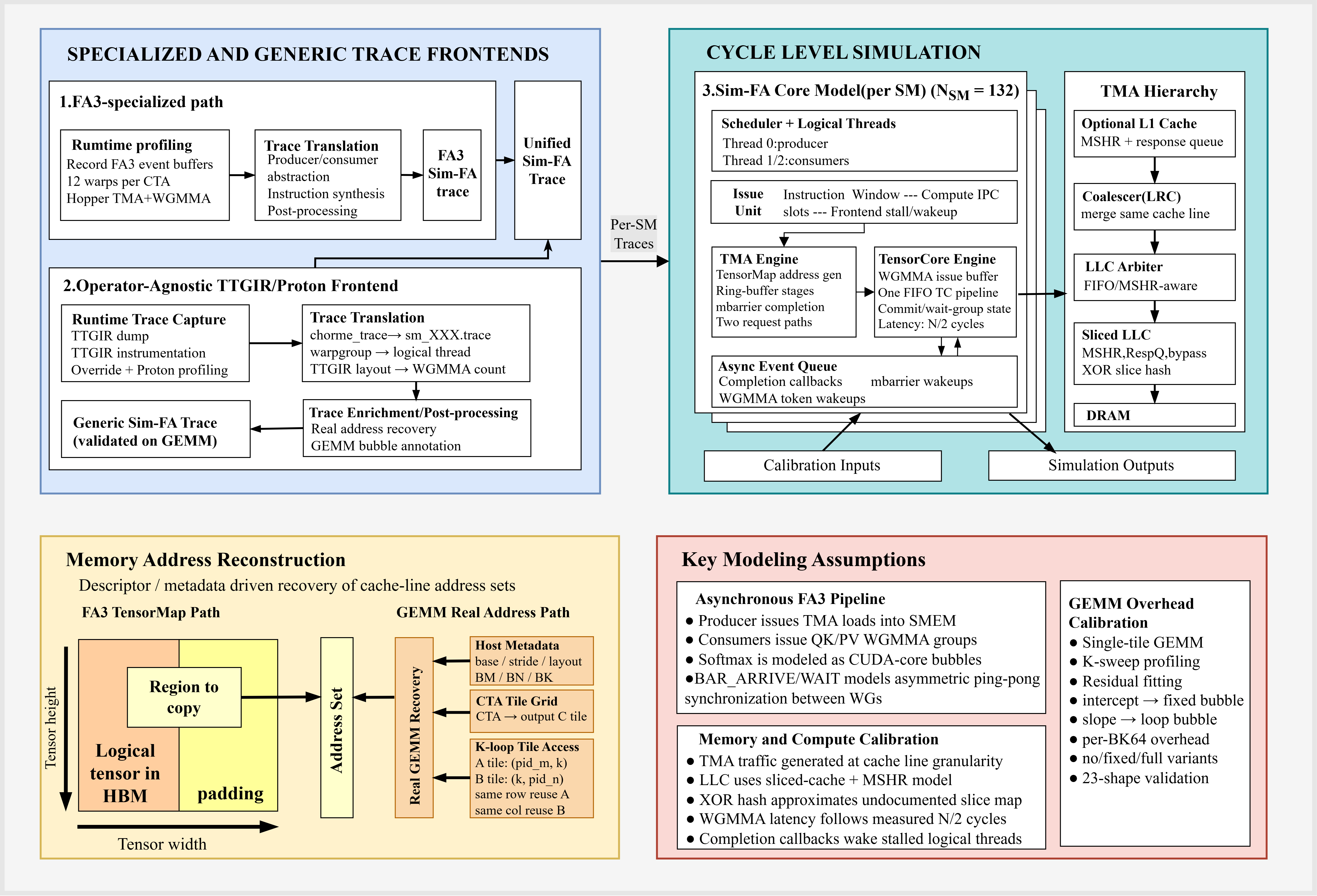}}
\vspace*{-1\baselineskip}
\caption{Sim-FA: Trace-Driven Cycle-Level Modeling of Hopper TMA/WGMMA Pipelines}
\label{fig:overview}
\vspace{-1\baselineskip}
\end{figure*}
\subsection{Design Scope and Abstractions}

\begin{table}[t]
  \centering
  \caption{Simulated System Configurations}
  \vspace*{-1\baselineskip}
  \label{table:simconfig}
  \scriptsize
\begin{tabular}{p{1.1cm}p{6.4cm}}
\hline
Base & 1830MHz, 132 SMs (66 TPCs), 50MB L2 cache, 80 L2 slices, 989 TFLOPS \\
\hline
SM & \begin{tabular}[t]{@{}p{6cm}@{}} WGMMA\_issue\_buffer\_size=16, MMA (Dense) precision=FP16 \end{tabular} \\
\hline
TMA Engine & (per SM) cache-lines-per-cycle=2, max-inflight-lines=64 \\
\hline
L2 slice & near-latency=258, far-latency=414, alloc-on-fill, write-back, write-allocate, resp\_q\_size=128, req\_q\_size=32 \\
\hline
Memory & HBM3-5200, 80 channels, using Ramulator2.0 \cite{ramulator2} \\
\hline
\end{tabular}
\vspace*{-1\baselineskip}
\end{table}

The new programming paradigms introduced in Section~\ref{advanced_feature} are different from the traditional SIMT programming model in several aspects. In the classic SIMT model, each thread in a block owns a data slice and synchronizes at block granularity. However, in the new programming model, a WarpGroup functions as a whole to perform computation and memory access collectively, and the barriers between the producers and consumers partially reduce the need for frequent block-level synchronization. Legacy cycle-accurate GPU simulators are tailored to SIMT programming models, making it difficult for them to support different types of asynchronous primitives.

To accommodate Hopper's asynchronous programming model, Sim-FA adopts a WarpGroup-oriented abstraction: each WarpGroup is modeled as a logical thread with a single instruction stream, becoming a physical thread once resident on an SM and consuming shared-memory and register resources. This abstraction is intentionally scoped to kernels whose performance is dominated by TMA/WGMMA execution and their associated synchronization and memory interactions; it does not aim to reproduce every instruction-level behavior of a general-purpose GPU. Instructions are classified into MMA, memory access, activation, and other computation, with activation latency modeled as cycle bubbles. We explicitly model asynchronous primitives and barriers because they determine the overlap among memory movement, matrix computation, and activation; the supported primitives are summarized in Table~\ref{tab:simfa-isa}.

Figure~\ref{fig:overview} summarizes the Sim-FA workflow. This section describes the simulator core and memory hierarchy; the generic TTGIR and FA3-specific frontends are presented in Sections~\ref{generic_frontend} and~\ref{fa3_e2e_pipe}. Sim-FA models only timing-visible structures critical to TMA/WGMMA performance, including TMA traffic, Tensor Core occupancy, barrier wakeups, and memory contention. All GEMM and FA3 experiments use the same H800-aligned configuration in Table~\ref{table:simconfig}, without operator-specific retuning.


  \begin{table*}[t]
  \centering
  \caption{Sim-FA instruction set.
    \textit{map}: Tensor Map ID;
    \textit{sid}: mbarrier/stage (ring-buffer slot);
    \textit{gid}: async group ID;
    \textit{N}: max outstanding groups;
    \textit{bid}: barrier ID;
    \textit{M,N,K}: WGMMA tile shape;
    \textit{cyc}: cycle count.}
    \vspace*{-1\baselineskip}
  \label{tab:simfa-isa}
  \small
  \setlength{\tabcolsep}{6pt}
  \renewcommand{\arraystretch}{1.22}
  \begin{tabular}{@{}p{1.7cm} p{4.2cm} p{4.6cm} p{6.4cm}@{}}
  \toprule
  \textbf{Category} & \textbf{Instruction} & \textbf{Hopper primitive} &
  \textbf{Semantics} \\
  \midrule
  Tensor Map
    & \texttt{DEF\_TMAP map rank dims strides box esz}
    & \texttt{cuTensorMapEncodeTiled}
    & Declare a Tensor Map descriptor for later TMA. \\
  \midrule
  \multirow{3}{*}{TMA (Load)}
    & \texttt{TMA\_TENSOR smem gmem map sid}
    & \texttt{cp.async.bulk.tensor}
    & Async HBM$\to$SMEM tile copy. \\
    & \texttt{MB\_WAIT sid}
    & \texttt{mbarrier.try\_wait}
    & Wait for TMA \textit{sid}. \\
    & \texttt{ACQUIRE\_STAGE} / \texttt{RELEASE\_STAGE sid}
    & \texttt{pipeline.acquire} / \texttt{release}
    & Producer waits for / consumer frees ring-buffer slot \textit{sid}. \\
  \midrule
  \multirow{2}{*}{TMA (Store)}
    & \texttt{TMA\_STORE smem gmem map gid}
    & \texttt{cp.async.bulk.tensor} (store)
    & Async tile store in group \textit{gid}. \\
    & \texttt{TMA\_COMMIT gid} / \texttt{TMA\_WAIT gid N}
    & \texttt{cp.async.bulk.commit\_group} / \texttt{wait\_group}
    & Seal group \textit{gid}; wait until $\le$ \textit{N} groups remain. \\
  \midrule
  \multirow{2}{*}{WGMMA}
    & \texttt{WGMMA d a b M N K gid}
    & \texttt{wgmma.mma\_async}
    & Issue async $M{\times}N{\times}K$ MMA into group \textit{gid}. \\
    & \texttt{WGMMA\_COMMIT gid} / \texttt{WGMMA\_WAIT gid N}
    & \texttt{wgmma.commit\_group} / \texttt{wait\_group}
    & Seal group \textit{gid}; wait until $\le$ \textit{N} groups in flight. \\
  \midrule
  \multirow{2}{*}{Ping-pong Sync}
    & \texttt{BAR\_ARRIVE bid}
    & \texttt{bar.arrive}
    & Non-blocking signal on barrier \textit{bid}. \\
    & \texttt{BAR\_WAIT bid k}
    & \texttt{bar.sync}
    & Wait for $\ge$ \textit{k} arrivals on \textit{bid}. \\
  \midrule
  Bubble
    & \texttt{bubbles cyc}
    & (CUDA cores)
    & Stall \textit{cyc} cycles for non-MMA work, e.g., softmax. \\
  \bottomrule
  \end{tabular}
  \end{table*}

\subsection{Event-Driven Execution}

\begin{algorithm}[t]
\caption{Per-cycle execution of a Sim-FA core}
\label{alg:simfa-core}
\small
\begin{algorithmic}[1]
\While{program is not complete}
\State $cycle \gets cycle + 1$
\State \textit{Scheduler:} Select a ready logical thread
\State Process completed async events and wake blocked threads
\State Tick TMA Engine and TensorCore Engine
\State Retire ready instructions from each physical thread

    \State Fetch next trace instruction
    \If{the instruction issues an asynchronous operation}
        \State Enqueue the task into the corresponding async engine
    \ElsIf{the instruction is wait and condition is unsatisfied}
        \State Roll back the logical PC and mark thread as stalled
        \State Register the thread in the corresponding waiter list
    \Else
        \State Insert ready instruction or consume compute bubbles
    \EndIf
\EndWhile
\end{algorithmic}
\end{algorithm}
As shown in Figure~\ref{fig:overview}, the Sim-FA core model has five main components. The \textit{Scheduler} checks whether there are available physical thread slots, considering the hardware resource constraints. If so, it will dispatch a new logical thread to the core. When all logical threads have been dispatched and completed, program execution terminates. The \textit{Scheduler} also functions as a "warp scheduler" in the simulator. It selects an available physical thread for the \textit{Issue Unit}, and only the selected physical thread can issue instructions. Many scheduling strategies are supported, such as Greedy-then-Oldest (GTO). This means the scheduler keeps selecting a physical thread until it stalls; then it selects the oldest one that is not stalled.
The \textit{Issue Unit} follows an in-order issue, out-of-order execution model. It issues instructions for the selected physical thread, dispatches special instructions to the corresponding units, such as the \textit{TMA Engine} and \textit{TensorCore Engine}, and manages the stalling and wakeup of physical threads. 

The \textit{TMA Engine} generates addresses from abstract TensorMap descriptors. It also manages mbarrier dependencies and pipeline dependencies from producers to consumers. 
The \textit{TensorCore Engine} consists of a single tensor core pipeline and an instruction buffer that manages the unfinished WGMMA instructions. For the FP16 cases used in this work, we model dense WGMMA as $m64nNk16$ with fixed $M=64$, $K=16$, and Hopper-supported $N$ values. Following the instruction-level microbenchmark results in~\cite{luo2025hopper}, we model the completion latency of FP16 WGMMA as approximately $N/2$ cycles for sufficiently large shapes.
It also manages the instruction group commits and dependencies required by the program. Shared stall-and-wakeup logic is implemented as a C++ class, the \textit{Asynchronous Event Queue (AEQ)} , which is used by both the \textit{TMA Engine} and the \textit{TensorCore Engine} to simplify implementation.

Algorithm~\ref{alg:simfa-core} outlines this per-cycle flow: completed asynchronous events wake blocked threads, the two engines advance one cycle, and the selected physical thread issues its next instruction---dispatched to an engine if asynchronous, rolled back and stalled if it is a wait whose condition (e.g., an \texttt{MB\_WAIT}) is unsatisfied, or inserted into the instruction window with its bubbles otherwise.

\subsection{Modeling the Tensor Memory Accelerator}
This subsection details how Sim-FA models and evaluates the Tensor Memory Accelerator (TMA). While Figure~\ref{fig:overview} presents the TMA engine as one asynchronous backend component, here we further expose the timing-relevant mechanisms inside this engine for latency and bandwidth evaluation. In Level-1 TMA mode, each TMA operation is issued as real memory traffic through the cycle-level L2 and DRAM hierarchy, and its completion time is determined by the resulting traffic timing. We calibrate three parts: setup overhead, partitioned-L2 behavior, and finite request-tracking resources.
\begin{figure}[t]
  \centering
  \includegraphics[width=0.95\linewidth]{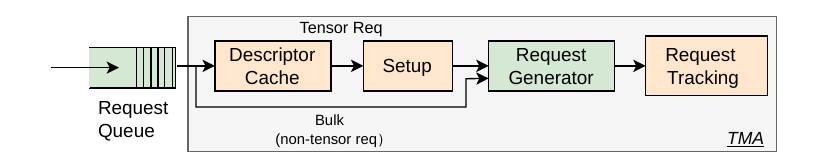}
  \vspace*{-1.1\baselineskip}
  \caption{TMA request path. Non-tensor bulk requests bypass the descriptor-cache and TensorMap setup path.}
  \label{fig:tma_bypassing}
\end{figure}

\textbf{\textit{Setup latency decomposition.}}
We separate TMA setup overhead from memory-system latency. Hopper microbenchmarks show TMA accesses are about 170 cycles slower than regular memory accesses, covering instruction issue, mbarrier initialization, and synchronization~\cite{luo2025hopper}. However, this overhead should not be uniformly applied to all TMA operations. Following the Tensor Memory Accelerator Unit (TMAU) patent~\cite{tma_patent}, Figure~\ref{fig:tma_bypassing} illustrates the two TMA request paths: non-tensor bulk requests bypass the descriptor cache and setup, directly entering request generation, while TensorMap requests first go through the descriptor cache and setup. We accordingly divide the setup overhead into a common launch latency (\textbf{40 cycles}) shared by all TMA operations and an additional TensorMap setup latency (\textbf{130 cycles}) charged only to TensorMap requests.

\textbf{\textit{Partitioned L2 and RemoteCopy proxy.}}
The Hopper L2 consists of two 25\,MB partitions linked by a high-speed channel~\cite{luo2025hopper}; far-partition accesses may incur cross-partition traffic. The random-pointer-chase benchmark of~\cite{luo2025hopper} reveals three regimes in the TMA latency curve: an L2-hit floor, a DRAM-bound plateau, and a wide fluctuating transition window between them, which a unified-LLC model cannot reproduce. Since Hopper's intra-L2 policy is undisclosed, Sim-FA does not attempt to reproduce a specific protocol and instead adds a behavioral \emph{remote-copy proxy} on top of the unified LLC. When a request from an SM resolves to a far slice, the LLC probabilistically inserts a shadow line into its near slice that competes with regular lines for capacity, mimicking the effect of cross-partition data movement in the real L2. The mechanism is controlled by two knobs, a maximum insertion probability and an occupancy threshold, both calibrated once against the H800 latency curve and held fixed across all subsequent experiments.

\begin{figure}[t]
  \centering
  \includegraphics[width=1.0\linewidth]{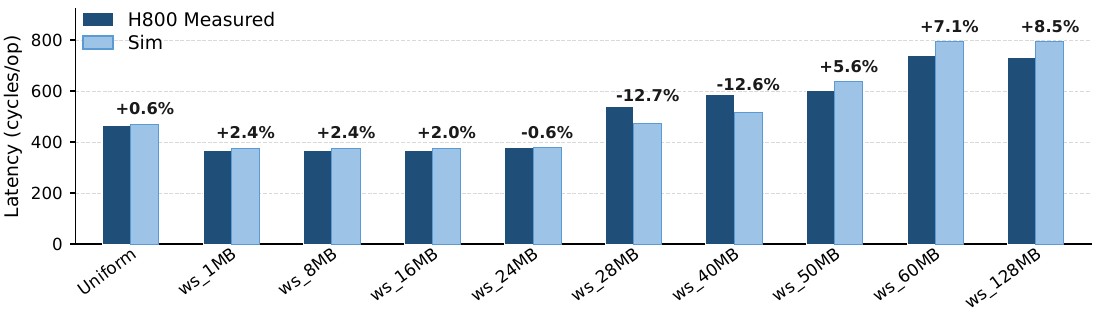}
  \vspace*{-1.5\baselineskip}
    \caption{TMA latency validation on H800. The RemoteCopy proxy is enabled for ws 25--50\,MB transition region; Sim-FA closely tracks the measured latency across all three regimes.}
  \label{fig:tma_latency}
\end{figure}

With the RemoteCopy proxy enabled across the partition-transition region, Figure~\ref{fig:tma_latency} compares Sim-FA against the measured H800 latency on 10 representative working sets covering the L2-hit floor, the fluctuating transition window, and the DRAM-bound plateau. The overall MAPE is \textbf{5.45\%} with a maximum absolute error of 12.73\% at ws\_28\,MB; excluding the 25--50\,MB transition region, MAPE drops to \textbf{3.65\%} with a worst case of 8.49\% at ws\_128\,MB.

\begin{figure}[t]
  \centering
  \includegraphics[width=1.0\linewidth]{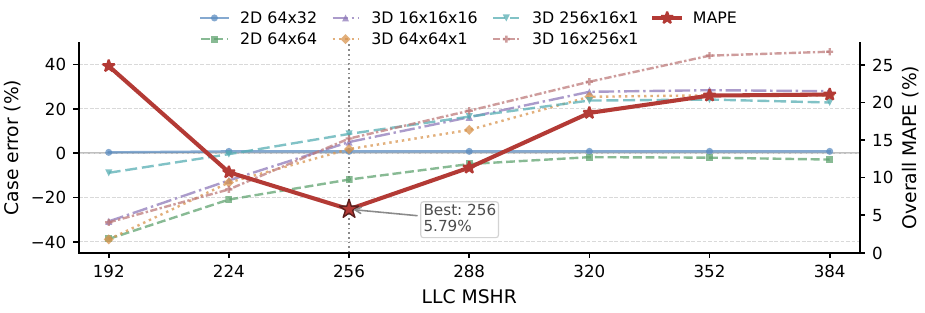}
  \vspace*{-1.5\baselineskip}
  \caption{MSHR sensitivity of TMA bandwidth cases. Finite request-tracking capacity is timing-visible for bursty TensorMap transfers; MSHR=256 minimizes overall MAPE at 5.79\%.}
  \label{fig:tma_mshr_sensitivity}
\end{figure}

\textbf{\textit{MSHR-limited bandwidth.}}
The MSHR pool is one of the finite request-tracking resources we calibrate and a key bandwidth limiter: once it fills, no new misses can be issued to DRAM, and the request path stalls. Although Sim-FA models the LLC MSHR pool explicitly, the actual MSHR count on Hopper is not publicly disclosed, so we sweep \texttt{llc\_num\_mshr} from 192 to 384 per slice on the large TMA tiles, where MSHR pressure is most visible. As Figure~\ref{fig:tma_mshr_sensitivity} shows, MSHR=256 sits at the inflection point: below it every bursty descriptor is throttled (MAPE 24.85\% at 192), above it the backpressure vanishes (MAPE 21.06\% at 384), and at 256 the overall MAPE is \textbf{5.79\%} with a worst-case absolute error of 11.96\% (2D 64$\times$64).

\begin{figure}[t]
  \centering
  \includegraphics[width=1.0\linewidth]{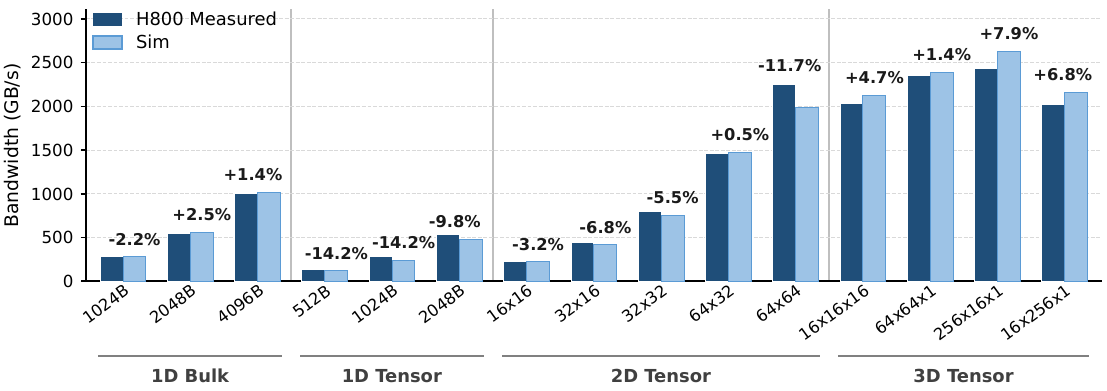}
  \vspace*{-1.5\baselineskip}
  \caption{TMA bandwidth validation on H800, covering bulk, 1D, 2D, and 3D TensorMap copies.}
  \label{fig:tma_bandwidth}
\end{figure}

Figure~\ref{fig:tma_bandwidth} evaluates TMA bandwidth using bulk, 1D, 2D, and 3D TensorMap copies. With MSHR=256, Sim-FA reaches an overall MAPE of \textbf{6.19\%} and a worst-case absolute error of 14.20\%. We attribute the residual error to microarchitectural features of the TMAU that Sim-FA does not model, in particular, its internal out-of-order issue and pipelining of address generation, and possibly shape-dependent dispatch strategies.

\section{Generic Frontend for TMA/WGMMA Kernels}
\label{generic_frontend}

To broaden Sim-FA's applicability beyond a single operator, we first build an operator-agnostic frontend that instruments and translates kernels at the TTGIR level. This general path stands in contrast to the FA3-specialized frontend we present next (Section~\ref{fa3_e2e_pipe}), which achieves high accuracy but is tightly coupled to FlashAttention-3's specific structure. 

\subsection{TTGIR-Based Trace Generation}
The generic frontend operates in four stages: TTGIR dump, instrumentation, override-and-profiling, and trace translation.

\subsubsection{Stage 1: TTGIR dump}
\noindent

The Triton compiler is invoked with \code{TRITON_KERNEL_DUMP=1}, which
emits TTGIR---an MLIR dialect lowered to GPU-specific operations. At this level, TMA loads, WGMMA instructions, and barriers
appear as TTGIR operations, including:
\begin{itemize}[leftmargin=1.5em]
  \item \code{ttng.async_tma_copy_global_to_local}
  \item \code{ttng.warp_group_dot}
  \item \code{ttng.wait_barrier}
  \item \code{ttng.arrive_barrier}
\end{itemize}

These operation names are defined by Triton's compiler internals and are independent of the high-level Python API (e.g.,Gluon).

\subsubsection{Stage 2: TTGIR Instrumentation}
\noindent

A source-to-source pass parses each dumped TTGIR file, matches TMA, WGMMA, and barrier operations, and inserts proton.record\_start / proton.record\_end markers around them. Each marker encodes a structured scope string containing the operation type, the tile dimensions $(M,N,K)$, the WGMMA instruction shape $(\mathit{instrM},\mathit{instrN},\mathit{instrK})$, the data types, the operand mode, and a site identifier. For example, a $128\times128$ WGMMA tile ($BK=64$), FP16 inputs and FP32 accumulation is recorded as:

\begin{verbatim}
SIMFA|WGMMA|M=128|N=128|K=64|instrM=64|
instrN=128|instrK=16|dtype=FP16|acc=FP32|
mode=RS|site=17
\end{verbatim}

\subsubsection{Stage 3: Override and Profiling}
\noindent

The Instrumented TTGIR is re-injected into Triton via 
\texttt{TRITON\_KERNEL\_OVERRIDE=1} and \texttt{TRITON\_OVERRIDE\_DIR}. Triton compiles the modified TTGIR and the Proton instrumentation records timestamps for every scope at runtime. The output is a \texttt{.chrome\_trace} file containing all TMA, WGMMA, and barrier events.

\subsubsection{Stage 4: Trace Translation}
\noindent

The translation module converts the \texttt{.chrome\_trace} into per-SM Sim-FA traces. The translator groups warps by WarpGroup($\mathit{warp\_id} / 4 \rightarrow \mathit{warpgroup\_id}$) and deduplicates identical events from the four warps within each WG. Each WG is then mapped to one Sim-FA logical thread, consistent with the 
abstraction of Section~\ref{core}.

Each WGMMA scope expands into
$\lceil M/\mathit{instrM} \rceil \times
 \lceil N/\mathit{instrN} \rceil \times
 \lceil K/\mathit{instrK} \rceil$
Sim-FA \texttt{WGMMA} instructions, where $\mathit{instrM}$ is derived from the WGMMA instruction shape and $\mathit{warpsPerCTA}$. For Hopper FP16, this yields WG-level $m64nNk16$ granularity; the example of Stage~2 therefore expands into $2\times1\times4=8$ instructions belonging to a single asynchronous group.
The timestamps recorded by Proton are used only to establish per-warp program order and to perform the deduplication above. After runtime collection, each Proton scope in the trace is translated into one or more Sim-FA instructions according to a fixed mapping. 

\begin{table}[t]
\centering
\caption{Translation from Proton trace scopes to SimFA instructions.}
\label{tab:scope-mapping}
\footnotesize
\setlength{\tabcolsep}{4pt}
\renewcommand{\arraystretch}{1.12}
\begin{tabular}{@{}p{0.30\columnwidth}p{0.62\columnwidth}@{}}
\hline
\textbf{Proton scope} & \textbf{Generated SimFA instruction} \\
\hline
\texttt{TMA\_LOAD}
&
\texttt{TMA\_TENSOR} with shared-memory address, global-memory
address, tensor map, and barrier identifier.
\\

\texttt{MB\_WAIT}
&
\texttt{MB\_WAIT} on the corresponding memory barrier.
\\

\texttt{WGMMA}
&
\texttt{WGMMA} followed by \texttt{WGMMA\_COMMIT} for the same
warp-group instruction group.
\\

\texttt{WGMMA\_WAIT}
&
\texttt{WGMMA\_WAIT} with the recovered group identifier and pending
count.
\\

\texttt{TMA\_STORE}
&
\texttt{TMA\_STORE} followed by \texttt{TMA\_COMMIT}.
\\

\texttt{TMA\_WAIT}
&
\texttt{TMA\_WAIT} with the recovered group identifier and pending
count.
\\
\hline
\end{tabular}
\end{table}

Table~\ref{tab:scope-mapping} summarizes the fixed translation table. Because instrumentation targets TTGIR ops rather than Python APIs, the same pipeline processes any kernel that uses TMA and WGMMA which is independent of the high-level DSL (e.g., Triton, Gluon).


\subsection{Validation on GEMM}
GEMM is the most fundamental operator in deep learning and an ideal validation target for the generic frontend. 

\subsubsection{GEMM Kernel Configuration}
The kernel is a Triton/Gluon matmul with TMA and WGMMA, without warp specialization (one WarpGroup of four warps per CTA), a $128\times128\times64$ tile, FP16 inputs with FP32 accumulation, and a grid of $\lceil M/128\rceil\times\lceil N/128\rceil$ CTAs; each K-loop iteration issues TMA loads of A and B, \texttt{MB\_WAIT}, WGMMA, and \texttt{WGMMA\_WAIT}, with a TMA store of the C tile in the epilogue. GEMM has no softmax or other non-MMA operations---its timing-visible work consists exclusively of TMA loads and WGMMA computation---making it a clean test of the frontend's core translation fidelity without confounding factors from operator-specific post-processing.

\subsubsection{Fair Timing}
A key challenge in simulation validation is isolating kernel execution from measurement artifacts. We use the CUPTI cold-L2 method: the L2 cache is flushed before every launch, and CUPTI extracts only the GPU kernel duration. This method matches the simulator's cold-cache, zero-host-overhead behavior by excluding host driver overhead, CUDA graph scheduling gaps and warm-cache effects.

\subsubsection{Bubble Calibration}
Even with fair timing, a gap still remains between measured and simulated cycles. This gap separates into two physically distinct components: (1) A K-independent prologue/epilogue overhead that the Proton scope stream does not cover (CTA admission, setmaxnreg, mbarrier initialization, TensorMap setup, store drain) . (2) A per-iteration latency mismatch from SimFA's idealized TMA and WGMMA $N/2$ models.

To separate them, we run a single-tile GEMM ($M {=} N {=} 128$, grid $1 {\times} 1$) with a K-sweep ($K = 64, 128,\ldots 32768$). One CTA on one SM can eliminate grid-launch and cross-CTA effects. Both real and simulated cycle counts are affine in $K$. Ordinary least squares fits:
\begin{equation}
    r(K) = \mathit{cyc}_{\mathit{real}}(K) - \mathit{cyc}_{\mathit{sim0}}(K) = \mathit{slope} \times K + \mathit{intercept}
\end{equation}

In this formula, the intercept is the fixed prologue/epilogue bubble. The slope is reported separately because a constant cannot correct a K-dependent error without distorting results at other K values.

\begin{figure}[t]
\centering
\includegraphics[width=\linewidth]{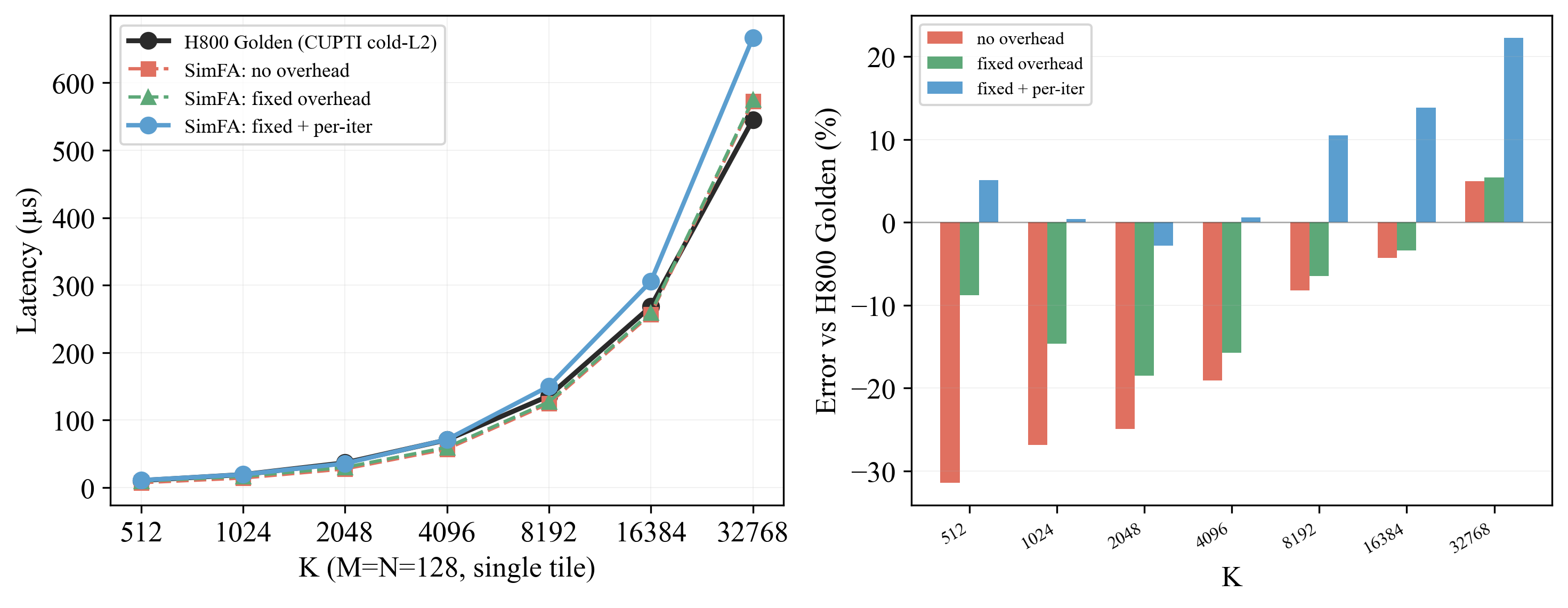}
\caption{K-sweep on single-tile GEMM ($M{=}N{=}128$).
Left: latency vs.\ $K$ for H800 golden and three SimFA variants.
Right: per-variant error with OLS decomposition into fixed bubble
(4314 cycles, shaded) and per-iteration residual.}
\label{fig:gemm-k-sweep}
\end{figure}

On H800 (1830MHz locked), this yields a fixed bubble of 4314 cycles (${\sim}2.36~\mu$s) and a per-iteration residual of 330 cycles per $BK{=}64$ iteration. The fit is performed on \(K = 512, 1024, 2048, 4096\) (\(R^2 = 0.975\)); the larger \(K\) values in the 23-shape set (\(8192, 16384, 32768\)) are held out from the calibration and serve as an extrapolation check. 

\subsubsection{23-Shape Results}
Table~\ref{tab:gemm-variants} reports MAPE across 23 GEMM shapes under three calibration levels.

\begin{table}[t]
\centering
\caption{MAPE across 23 GEMM shapes under progressive calibration.}
\vspace*{-0.8\baselineskip}
\label{tab:gemm-variants}
\footnotesize
\setlength{\tabcolsep}{3pt}
\renewcommand{\arraystretch}{1.08}
\begin{tabular}{@{}lccc@{}}
\toprule
\textbf{Variant} & \textbf{MAPE} &
\shortstack{\textbf{Within}\\\(\pm 5\%\)} &
\shortstack{\textbf{Within}\\\(\pm 10\%\)} \\
\midrule
No overhead             & 16.79\% & 3 / 23  & 4 / 23  \\
Fixed overhead only       & 11.90\% & 3 / 23  & 10 / 23 \\
Fixed + per-iter residual & 5.49\%  & 12 / 23 & 19 / 23 \\
\bottomrule
\end{tabular}
\end{table}

\begin{figure}[t]
\centering
\includegraphics[width=\linewidth]{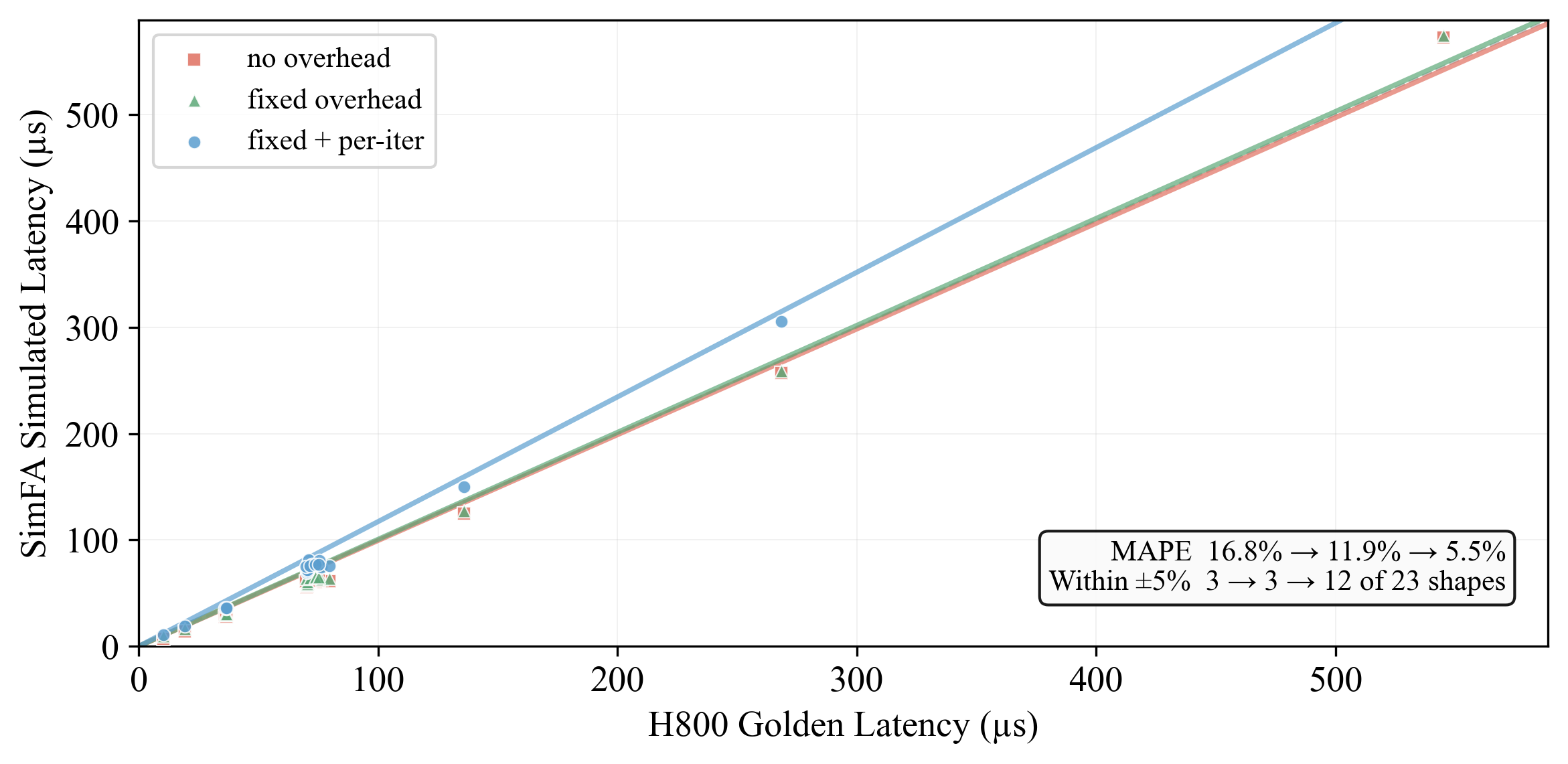}
\caption{Simulated vs. measured latency for 23 GEMM shapes under three calibration variants.}
\label{fig:gemm-parity}
\end{figure}

Without any bubble, the simulator systematically underestimates latency. Adding the fixed launch overhead (4314 cycles) improves small-K shapes where prologue/epilogue dominates. And adding the per-iteration residual (330 cycles / BK64) reduces MAPE to 5.49\%, with 12 of 23 shapes within \(\pm\)5\% of H800 golden measurements. The calibration involves only two parameters, obtained by an ordinary-least-squares fit on a single-tile microbenchmark and then held fixed across all 23 held-out shapes. The resulting 5.49\% MAPE is comparable to the 5.7\% MAPE of the FA3-specialized path (Section~\ref{fa3_e2e_pipe}), indicating that the two frontends reach similar fidelity levels on their respective operators.
Both the calibration sweep and the 23-shape evaluation are run with the real-address recovery of Section~\ref{sec:realaddr} enabled, so the residual is not an artifact of synthetic addressing. The two constants belong to the GEMM frontend rather than to the core: the intercept covers prologue/epilogue work that the Proton scope stream does not record, while the slope absorbs the TMAU-internal issue pipeline that Sim-FA does not charge and the optimism of the $N/2$ WGMMA law inside a tight K-loop (Section~\ref{scope}). 
Neither constant is applied to the FA3 path of Section~\ref{fa3_e2e_pipe}, whose non-MMA time is modeled explicitly as bubbles.

\subsection{Real Address Recovery}
\label{sec:realaddr}
By default, the generic translator derives each CTA's address from the tile shape alone, ignoring the tensor's true global-memory layout. This collapses each tile's hardware-aligned row stride into a contiguous block, so a tile spanning many cache lines on real hardware is compressed into a few, drastically undercounting DRAM cache-line fetches. A secondary opposite effect---CTAs in the same grid row receiving disjoint synthetic addresses for the same tile, hiding shared data from the L2---inflates traffic, but is far smaller and masked by the first. The net result is that simulated DRAM traffic is underestimated 10--30\(\times\).

We address this with a \emph{real-address recovery} mechanism. Before launch, a host-side pass captures each tensor's base pointer, shape, byte strides, block shape, and element size, and reconstructs the true global-memory address of every tile access. For tiled GEMM, per-CTA coordinates follow an affine mapping: \(A\) at \((pid\_m, k)\), \(B\) at \((k, pid\_n)\), \(C\) at \((pid\_m, pid\_n)\). The real address is:

\[
\mathit{gmem} = \mathit{base} + pid\_m \times \mathit{stride}_m + pid\_n \times \mathit{stride}_n + k \times \mathit{stride}_k
\]

After recovery, CTAs sharing the same A row-block (same \(pid\_m\)) emit identical \(\mathit{gmem}\) addresses, enabling L2 cross-CTA reuse detection. DRAM read traffic rises from 0.03--0.09\(\times\) of the analytic footprint to 1.2--1.5\(\times\). 



\section{Simulating FlashAttention-3 End-to-End}
\label{fa3_e2e_pipe}
    
Unlike the Triton kernels of Section~\ref{generic_frontend}, the reference FA3 implementation is written in CUTLASS and offers no compiler IR to instrument. We therefore instrument the kernel source directly, and recover the operator-specific structure that the generic translator cannot express.
The FlashAttention-3 kernel relies heavily on asynchronous primitives such as TMA, WGMMA, and mbarrier. Its implementation divides each CTA into one producer WarpGroup and two consumer WarpGroups, challenging the traditional SIMT-based GPGPU simulators. Benefiting from the wide support of these asynchronous primitives introduced in Section~\ref{async_pipe}, our simulator can model the asynchronous pipeline of FlashAttention-3. However, the input format of the simulator is traces with abstract operation codes. We need to convert FA3 kernel events into the trace format supported by the simulator.
 
Figure~\ref{fig:overview} shows the end-to-end pipeline, which comprises three phases: kernel instrumentation, offline trace translation, and simulation. The design follows three principles: minimal disturbance during profiling, offline rather than runtime interpretation of recorded events, and simulation parameters derived from hardware configurations and microbenchmarks.
\subsection{Runtime Instrumentation in the FlashAttention-3 Kernel }
Lane~0 of each warp records nine event types in a lock-free per-warp buffer. On the producer side, \code{TMA_Q_ISSUE}, \code{TMA_K_ISSUE}, and \code{TMA_V_ISSUE} are emitted before the corresponding \code{cp.async.bulk.tensor} instructions. On the consumer side, \code{WAIT_K_START}/\code{WAIT_V_START} are recorded at \code{mbarrier.try_wait} entry, \code{GEMM_QK_ISSUE}/\code{GEMM_PV_ISSUE} before the corresponding \code{wgmma.mma_async} instructions, and \code{WAIT_WG_1}/\code{WAIT_WG_0} at \code{wgmma.wait_group} 1/0.


\subsection{Offline Trace Translation}
Since profiling is recorded per warp whereas the simulator treats the WarpGroup (four warps) as a whole (Section~\ref{async_pipe}), the offline pass first deduplicates the events, yielding logs from three WarpGroups: \textit{Producer}, \textit{Consumer 1}, and \textit{Consumer 2}, corresponding to the three logical threads in the simulator.

The recorded events are then translated into instructions systematically. When \texttt{TMA\_K\_ISSUE} occurs, we translate the event into two simulator instructions: \texttt{ACQUIRE\_STAGE} and \texttt{TMA\_TENSOR}. The former expresses the dependency from producer to consumer, and the latter is the actual memory access instruction. The K tile index, rather than the full address, is recorded in the event payload, and the complete address is reconstructed using the tensor base address, tile size, and stride information given in the TensorMap definition. The V tiles are handled in the same way. The event \texttt{GEMM\_QK\_ISSUE} is expanded into eight WGMMA instructions with the same group id (\emph{gid}), following the same $\lceil M/\mathit{instrM}\rceil \times \lceil N/\mathit{instrN}\rceil\times \lceil K/\mathit{instrK}\rceil$ expansion rule as Section~\ref{generic_frontend}-A. The event \texttt{GEMM\_PV\_ISSUE} is handled similarly. \texttt{WAIT\_WG\_1} and \texttt{WAIT\_WG\_0} are translated into \texttt{WGMMA\_WAIT gid 1} and \texttt{WGMMA\_WAIT gid 0}, respectively.

FlashAttention-3 uses ping-pong scheduling to overlap MMA with softmax. When \textit{Consumer 1} performs MMA, \textit{Consumer 2} executes softmax, and vice versa. As in the original kernel, we model the mechanism by inserting \texttt{BAR\_WAIT} and \texttt{BAR\_ARRIVE} into the traces. For \textit{Consumer 1}, the trace emits \texttt{BAR\_ARRIVE} before QK MMA, and emits \texttt{BAR\_WAIT} before softmax, and vice versa. The \texttt{RELEASE\_STAGE} for the K tile is inserted right after \texttt{WAIT\_WG\_1} because the QK MMA has finished. The \texttt{RELEASE\_STAGE} for the V tile is inserted after \texttt{WAIT\_WG\_0}.

Non-MMA operations, including softmax, rowmax, and rowsum, are modeled as a number of bubbles after \texttt{WAIT\_WG\_1}. For each output tile processed by each consumer WarpGroup, we estimate the cycles taken by these operations as follows:

\begin{itemize}[leftmargin=0.5cm]
    \item $\text{rowmax}(\mathbf{S}_i^{(j)})$: $T_M\cdot T_N=64\times176=11264$ comparisons, while there are $4\times32=128$ threads in the WarpGroup that can perform 1 FP32 comparison per cycle $\rightarrow 88$ cycles;
    \item $\widetilde{\mathbf{P}}_i^{(j)}=\exp(\mathbf{S}_i^{(j)}-m_i)$: $11264$ exponential operations, H800 MUFU throughput: 16 ops/SM/cycle \cite{fa3} $\rightarrow 704$ cycles;
    \item $\text{rowsum}( \widetilde{\mathbf{P}}_i^{(j)})$: $11264$ ops, 128 ops/cycle (FP32) $\rightarrow 88$ cycles;
    \item $\widetilde{\mathbf{P}}_i^{(j)}$ FP32 to FP16: 11264 ops, 256 ops/cycle (FP16) $\rightarrow 44$ cycles;
    \item rescale $\mathbf{O}_i$: $T_M\cdot D=64\times 128=8192$ multiply-adds, 128 ops/cycle (FP16) $\rightarrow 64$ cycles;
\end{itemize}

Similar to $T_M$, here $T_N$ denotes the tile size of the sequence length of KV tensors. $T_M=64$ and $T_N=176$ align with the kernels from which we extract traces. Tensor notation follows the definition in the FlashAttention-3 paper \cite{fa3}. Here, we do not include scalar operations since their contribution is orders of magnitude smaller. Summing up the above numbers yields a total of 988 cycles.

\subsection{End-to-End Validation}
\begin{figure}[!htbp]
    \centering
    \includegraphics[width=\linewidth]{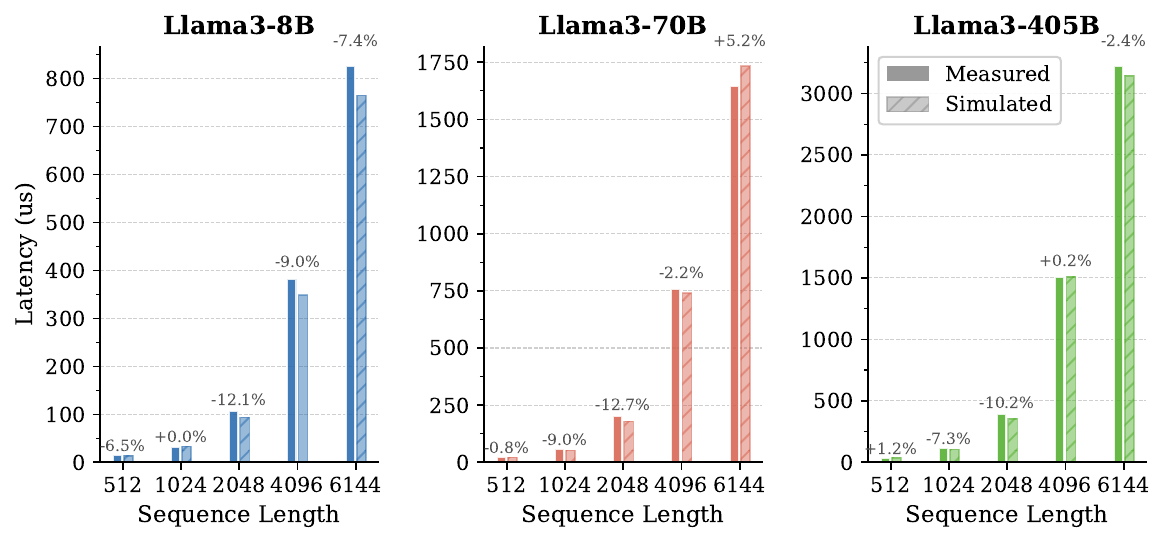}
    \vspace*{-1.5\baselineskip}
    
    \caption{FA3 Kernel Latency: Measured vs. Simulated  (MAPE = 5.7\%)}
    \label{fig:fa3_latency}
\end{figure}

In order to verify our simulator across a wide range of scenarios, we evaluate three models (Llama 3 8B, 70B, and 405B) with five sequence length configurations (512, 1024, 2048, 4096, 6144). The measured latency of the FlashAttention-3 kernel is collected using NVIDIA Nsight Systems, and we run the original kernel \textbf{without any modifications} (such as instrumentation) to avoid instrumentation overhead. The instrumented code is used only to generate traces for the simulator. The frequency of the H800 GPU is locked at 1830 MHz, and the simulator is configured with the same frequency. The full comparison between measured and simulated latency is shown in Figure~\ref{fig:fa3_latency}. The simulator achieves a mean absolute percentage error (MAPE) of 5.7\% and a maximum absolute percentage error of 12.7\%. 

We also extended the experiments with an H20 configuration (78 SMs, 96-channel HBM3, and a WGMMA latency multiplier $k_{\mathrm{eff}}=5.176$, measured independently by an H20 WGMMA microbenchmark as the product of the 3.95 peak-throughput ratio and a per-instruction slowdown of 1.31 uniform across workloads) and performed cycle-accurate simulation of 15 FlashAttention-3 configurations. The simulated kernel times achieve a
mean absolute percentage error (MAPE) of \textbf{3.28\%}, with a maximum error of \textbf{5.4\%} at $h{=}32, s{=}512$, where fixed overheads dominate. Leave-one-out cross-validation (out-of-sample MAPE 3.29\%) confirms generalizability without overfitting.
 
\begin{figure}[!htbp]
    \centering
    \includegraphics[width=1\linewidth]{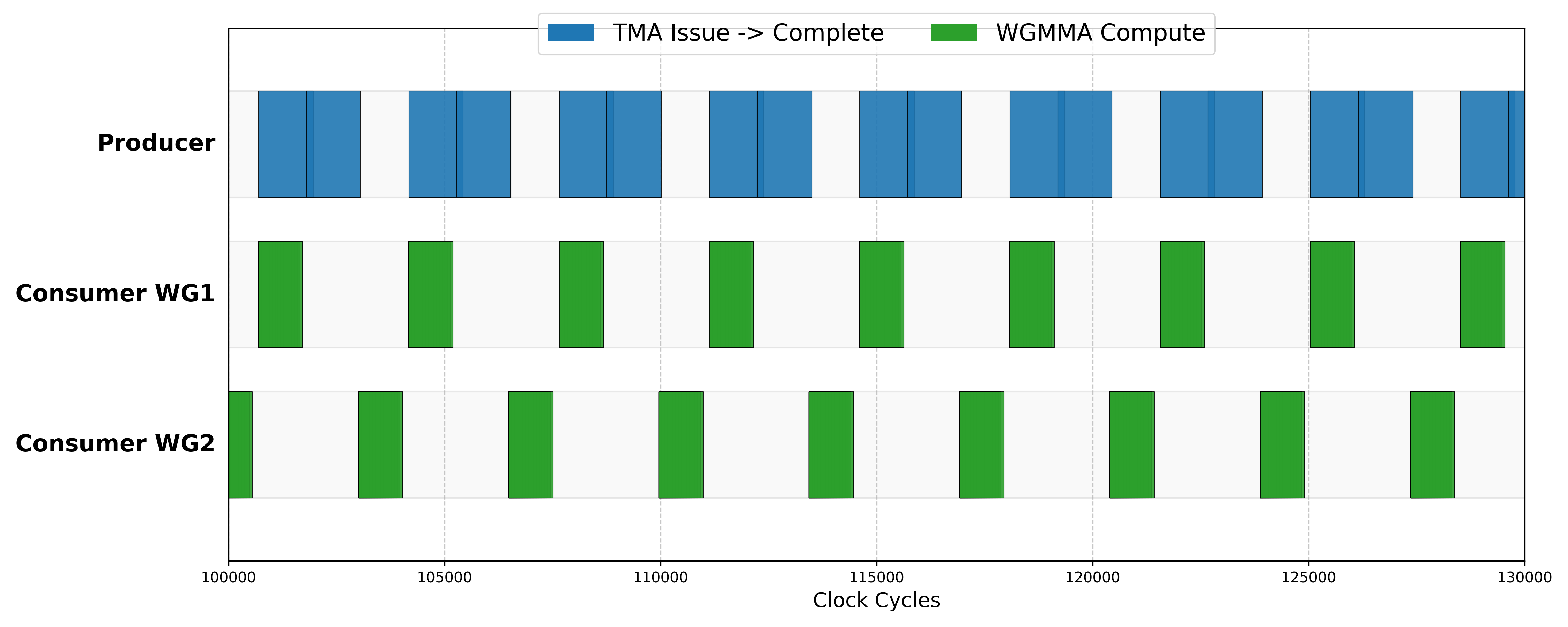}
    \vspace*{-1.5\baselineskip}
    
    \caption{FlashAttention-3 Pipeline Gantt Chart. (Llama-3 405B, sequence length = 6K, Streaming Multiprocessor ID 0).}
    \label{fig:fa3_gantt}
\end{figure}

We also use data collected from the simulation to draw the FlashAttention-3 pipeline Gantt chart (Figure~\ref{fig:fa3_gantt}). The chart clearly shows the timing overlap between the producer and two consumers, as well as the ping-pong scheduling between the two consumers to maximize tensor core utilization.

\subsection{Modeling Insights}
\label{modeling_insights}
During simulator tuning, many factors have been discovered that significantly influence performance. The analysis of such factors can offer insights into chip design. The ablation study results are shown in Table~\ref{tab:addr-hash-ablation}.

\begin{table}[t]
\centering
\caption{Ablation study results (MAPE)}
\vspace*{-1\baselineskip}
\label{tab:addr-hash-ablation}
\small
\begin{tabular}{lr}
\toprule
Variant & MAPE \\
\midrule
Sim-FA & 5.7\% \\
No LRC & 16.8\% \\
Oversimplified Hash Function & 64.3\% \\
No line deduplication & 511.4\% \\
\bottomrule
\end{tabular}
\end{table}
 
\textit{L2 Cache Request Coalescer (LRC):} NVIDIA Hopper GPUs provide an L2 Cache Request Coalescer that merges duplicate requests within a cluster before they reach the L2 cache. FlashAttention-3 manually coalesces requests across every pair of Streaming Multiprocessors (SMs), and we add coalescing units for every pair of SMs accordingly. Without modeling it, the MAPE would increase to 16.8\%.
 
\textit{Multi-Sliced Cache Allocation (Hash Function):}
Initially, we simply use the last few bits of the cache index to determine which cache slice the address belongs to. However, FlashAttention-3 generates highly regular memory accesses with certain strides, such as 2048 bytes. Such a simple allocation scheme can cause multiple requests to concentrate in certain cache slices and underutilize the cache bandwidth. Therefore, we use an XOR-based hash function ($slice = (line \oplus (line \gg 5)) \bmod N_{slice}$) to mitigate this imbalance. The MAPE would increase to 64.3\% without the XOR-based hash function due to performance deterioration.

\textit{TMA address generation (deduplication):} A TensorMap describes a logical tensor tile, and multiple elements in a tile may map to the same cache line. If we generate requests for each element, many duplicate requests will be generated, wasting system bandwidth. Therefore, duplicate requests should be removed in TMA units. Without request deduplication, the MAPE would increase to 511.4\%.

\section{Simulation Overhead}
\label{sim_characteristic}
A cycle-level GPU simulator inevitably incurs a significant wall-clock slowdown relative to real hardware. Sim-FA sustains 4.0--5.8~K simulated GPU cycles per wall-clock second across FA3 configurations; wall-clock time ranges from 5.3~s (h32, \(S{=}512\); 24.6~K cycles) to 285.7~s (h64, \(S{=}4096\); 1.36~M cycles). 
For LLM workloads, simulation wall-clock time is a first-order constraint: one transformer layer executes many TMA/WGMMA kernels; sweeps over model, batch, and sequence size, multiplied again by design-space exploration; and host memory bounds how many can run in parallel. We compare Sim-FA with Accel-Sim 2.0~\cite{accelsim2, AccelSim, accelsim0} on 23 operator-equivalent Hopper GEMM workloads. Both simulators are driven by the same per-CTA program---each iteration issues an mbarrier arrive/expect-tx pair, two TMA loads, eight WGMMA instructions, and the pipeline synchronization of Section~\ref{generic_frontend}. Both simulators run serially on the same ASUS GX10 host, three repetitions per configuration; the median is reported.

Sim-FA is faster in 22 of the 23 configurations with a 7.4$\times$ geometric-mean speedup (total 559.8~s vs.\ 106.2~s, i.e., 5.3$\times$). The largest gap is the 1024$\times$1024$\times$4096 GEMM: 178.6~s for Accel-Sim 2.0 versus 12.7~s for Sim-FA ($14.0\times$); the only exception is 2048$\times$128$\times$4096 at $0.8\times$. Peak host memory is 47--525~MB for Sim-FA versus 1.1--3.5~GB, a 5--23$\times$ reduction. 

Because both simulators are driven by the same hand-written per-CTA program rather than by a trace of the production kernel, this setup is a controlled \emph{cost} harness, and we do not use it to claim an accuracy advantage. The program omits address arithmetic, epilogue work, and polling loops that an instruction-level model must execute, and it carries neither the real-address recovery of Section~\ref{sec:realaddr} nor the frontend calibration of Section~\ref{generic_frontend}; under it both simulators underestimate CUPTI-measured H800 latency by a wide margin (68.2\% MAPE for Accel-Sim~2.0, 58.8\% for Sim-FA). The accuracy we claim for Sim-FA is the 5.49\% and 5.7\% obtained through its calibrated frontends (Sections~\ref{generic_frontend} and~\ref{fa3_e2e_pipe}); the comparison drawn here is confined to simulation cost and host memory. One structural reason for the cost gap concerns synchronization: the dynamic polling loops of these kernels are not static instruction artifacts and are distorted when replayed as such~\cite{accelsim2}. Sim-FA avoids this by giving \texttt{MB\_WAIT} and \texttt{WGMMA\_WAIT} blocking semantics at WarpGroup granularity, so a wait costs one stalled logical thread rather than a replayed spin-instruction stream.

\section{Analytical Model Validation}
\label{analytical_exp}
In this section, we validate SimFA-python, our analytical model framework, against ground-truth data collected via NVIDIA Nsight Compute (NCU) and demonstrate the pitfalls of traditional analytical models such as GenZ~\cite{genz}.

\subsection{Experimental Setup and Methodology}
\label{sec:setup}



All measurements are conducted on NVIDIA GB10 (Blackwell) and H800 (Hopper) GPUs. For GB10, we use a FlashAttention kernel implemented via cuTile, a domain-specific language (DSL); for H800, the official FlashAttention-3 implementation. Ground-truth latency and per-level memory traffic are collected with NCU in full-profile mode.

We adopt the Llama~3 family for its GQA shape parameters
(Table~\ref{tab:workloads}) rather than as an end-to-end deployment
scenario. All kernels are executed in the non-causal configuration on
hardware and modeled in the same configuration, so measured and
predicted quantities correspond to identical computations.

\subsection{Evaluation of Analytical Model}

\subsubsection{Validation of Last-Level Cache (LLC) Traffic}

\begin{figure}
    \centering
    \includegraphics[width=0.795\linewidth]{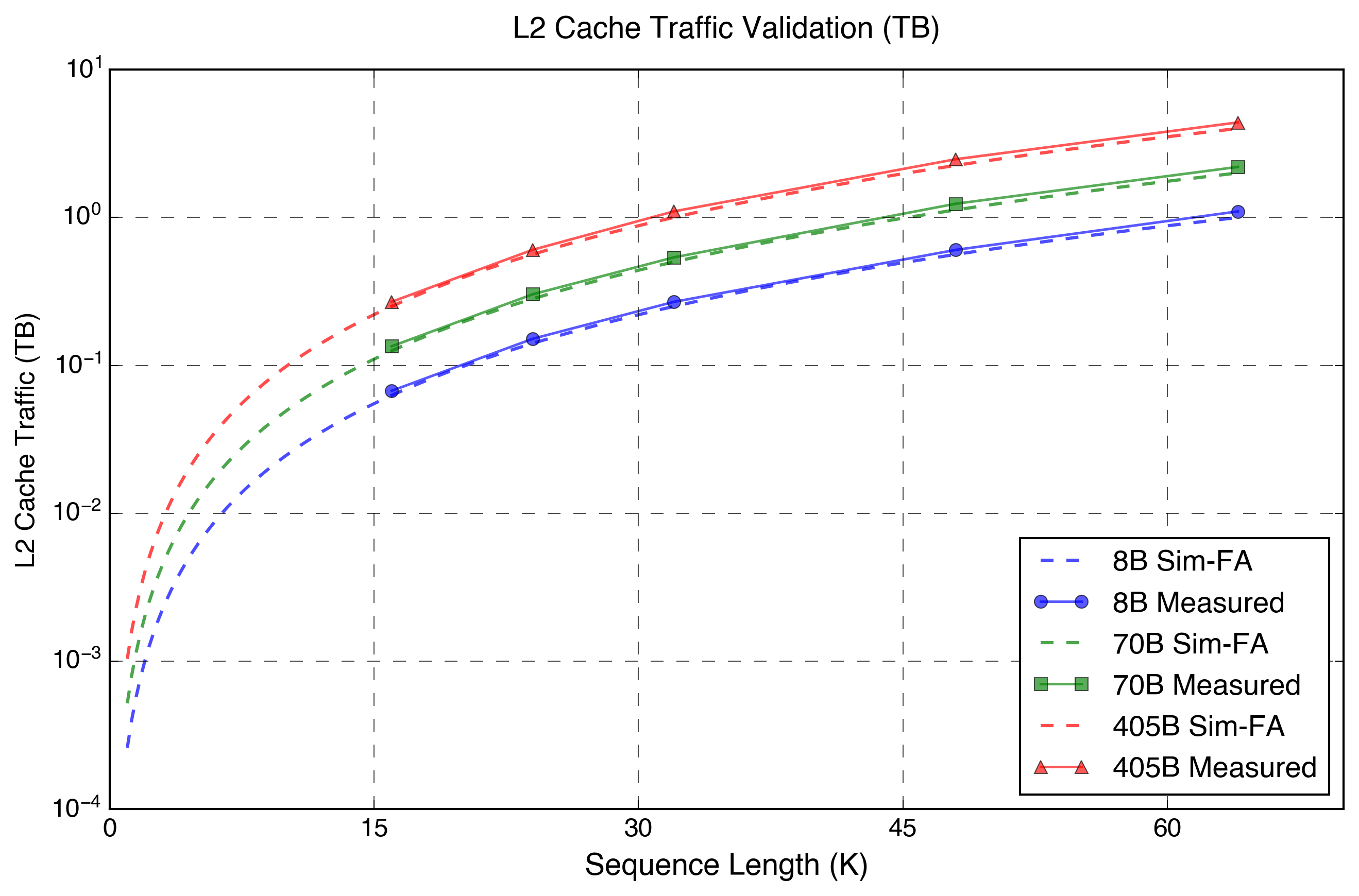}
    \vspace*{-1\baselineskip}
    \caption{L2 Cache Validation of SimFA-python on GB10.}
    \label{fig:simfapy_valida}
\end{figure}

We first evaluate whether SimFA-python can correctly estimate the
traffic observed at the last-level cache (i.e., L2 on these GPUs). This metric differs from
DRAM traffic: it reflects the total amount of memory requests generated
by the FlashAttention tiling schedule, before the requests are reduced by
cache reuse and request coalescing in the memory hierarchy. Accurately modeling LLC traffic is, therefore, a
prerequisite for analyzing the gap between ideal and realistic DRAM
traffic.

As derived in Eq.~(\ref{eq:l2_est}), each thread block loads one $Q$ tile,
stores one $O$ tile, and repeatedly traverses the $K$ and $V$ tensors for
the corresponding KV head. When the sequence length is sufficiently
large, the $K/V$ traversal dominates the total traffic, and the asymptotic
complexity becomes $O(L\cdot S/T_M)$. This term directly reflects the
number of query tiles and the amount of $K/V$ data scanned by each
tile.

Figure~\ref{fig:simfapy_valida} shows the comparison between
SimFA-python and NCU measurements for Llama 3
8B/70B/405B. The predicted LLC traffic closely matches the measured
results across different configurations. More
importantly, the measured curves follow the same scaling trend as the
analytical model, especially in the long-sequence regime where the
$K/V$ traffic dominates. This suggests that the proposed
thread-block-level formulation captures the major source of LLC
memory demand in FlashAttention.


\begin{table}[h]
    \caption{Workload Configurations based on Llama 3 Family}
    \vspace*{-1\baselineskip}
    \label{tab:workloads}
    \centering
    \begin{tabular}{l c c c c c}
        \toprule
        \textbf{Model} & \textbf{Hidden Size} & \textbf{$H_Q$} & \textbf{$H_{KV}$} & \textbf{$G$} & \textbf{$D$} \\
        \midrule
        Llama 3 \textbf{8B}   & 4,096  & 32  & 8 & 4  & 128 \\
        Llama 3 \textbf{70B}  & 8,192  & 64  & 8 & 8  & 128 \\
        Llama 3.1 \textbf{405B} & 16,384 & 128 & 8 & 16 & 128 \\
        \bottomrule
    \end{tabular}
\end{table}

\subsubsection{DRAM Traffic} 

We next validate the DRAM traffic model. Unlike LLC traffic, DRAM
traffic depends critically on whether the LLC can retain the reusable
$K/V$ working set. As discussed in Section~\ref{ideal} and
Eq.~(\ref{eq:cond}), the ideal-cache assumption holds only when this
working set fits in the effective LLC capacity.

SimFA-python therefore models DRAM traffic with two regimes. In the
\textit{ideal} regime, $Q$, $K$, and $V$ are read once from DRAM, and $O$ is written once, following Eq.~(\ref{eq:ideal}). In the
\textit{realistic} regime, the $K/V$ data may be re-fetched by multiple
waves of thread blocks, as modeled by Eq.~(\ref{eq:dram_real}). For long
sequences, this refetching term dominates, causing DRAM traffic to scale
as $O(L\cdot S)$, or $O(L^2)$ when $L=S$.

\begin{figure}
    \centering
    \includegraphics[width=0.99\linewidth]{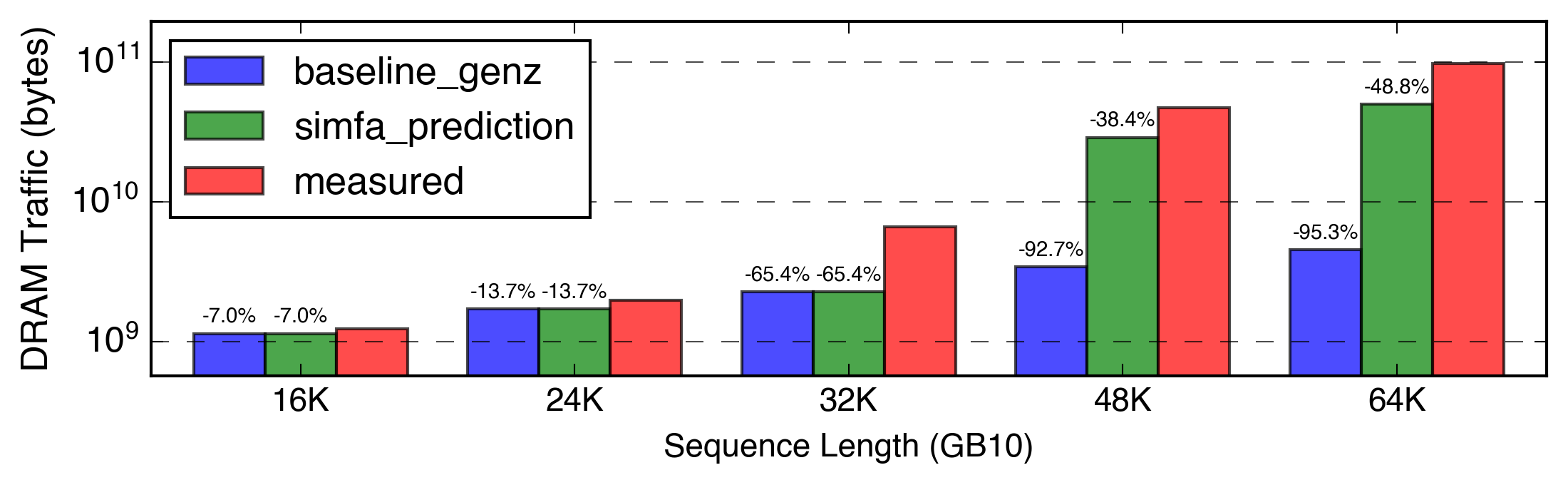}
    \includegraphics[width=0.99\linewidth]{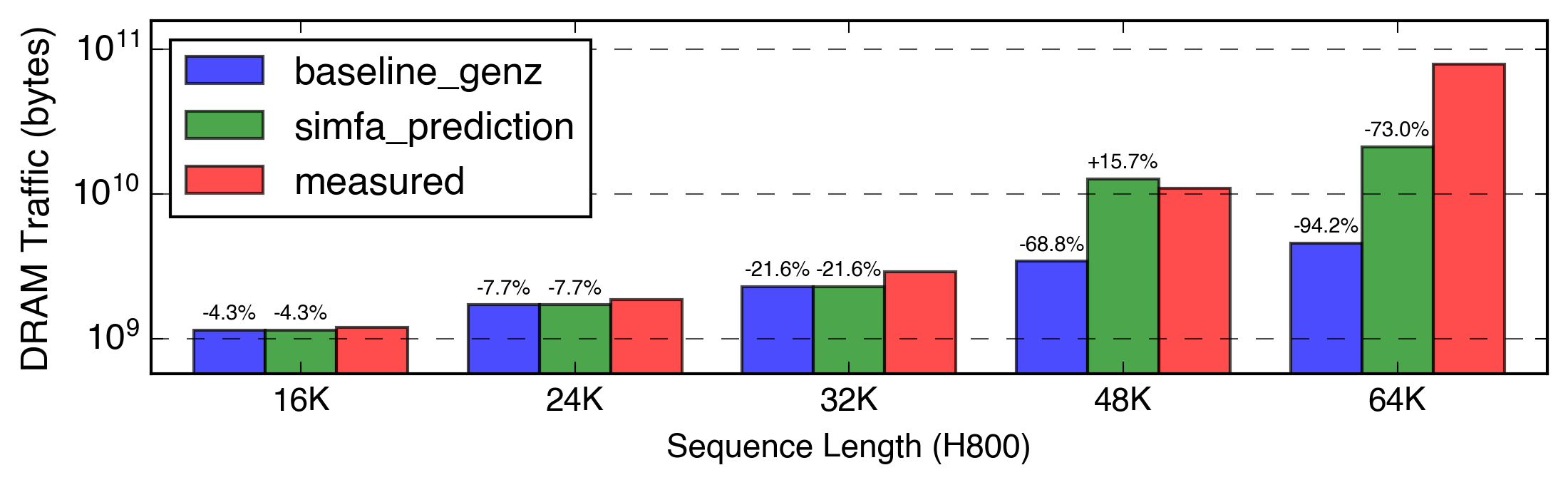}
    \vspace*{-1.2\baselineskip}
    \caption{DRAM traffic comparison of SimFA-python and baseline GenZ~\cite{genz} (Llama 3 405B, NVIDIA GB10 and H800).}
    \label{fig:dram_comp_genz}
\end{figure}

Figure~\ref{fig:dram_comp_genz} compares DRAM traffic predicted by
SimFA-python and GenZ~\cite{genz} with profiling results on NVIDIA GB10 and H800 GPUs. SimFA-python uses the ideal model when the ideal-cache condition holds, and switches to the realistic model otherwise. For H800, we use 25~MB, half of the nominal L2 capacity, as the effective cache
boundary, following the cache-partition behavior discussed in Section~\ref{core}.

At short sequence lengths, such as 16K and 24K, the measured DRAM traffic
is close to the ideal-regime prediction on both GPUs. This indicates that
the reusable $K/V$ working set can still be retained, merged, or otherwise
served efficiently by the cache hierarchy. As the sequence length grows,
the measured traffic increases more rapidly and begins to deviate from
the ideal model. This transition is especially visible beyond 32K--48K,
where the effective LLC capacity is no longer sufficient and repeated
$K/V$ refetching occurs across different waves of thread blocks.

At larger sequence lengths, especially 48K and 64K, the measured traffic
may further exceed the realistic-regime prediction. We attribute this gap
to inter-SM timing effects that are not fully captured by the analytical
model. Scheduling, synchronization, and pipeline-progress variations can
make memory requests arrive at different times, reducing request merging
and cache reuse opportunities. Such arrival order is a property of the executed schedule rather than of the shape parameters, and is therefore beyond the reach of any static formulation—it is precisely what the cycle-level core resolves.

\subsubsection{Comparison against Baseline Models}

The same results show why ideal-cache analytical models can mislead: GenZ~\cite{genz}, which does not model capacity-induced $K/V$ refetching, slightly overestimates DRAM traffic at short sequences (where reuse and coalescing are still effective) but significantly underestimates it at long sequences, where the $K/V$ working set exceeds the effective LLC capacity and is repeatedly refetched by multiple waves of thread blocks.


In contrast, SimFA-python captures this transition through the regime split and the concurrency-aware wave factor. This explains why prior ideal-cache-based models fail at large sequence lengths, where capacity-induced $K/V$ refetching dominates DRAM traffic and can mislead long-context attention design-space exploration.

\section{Scope and Limitations}
\label{scope}

Sim-FA retains only state that is timing-visible to an asynchronous pipeline.
That decision buys the speedup of Section~\ref{sim_characteristic} and
equally bounds what the model can answer. It is accurate while three
conditions hold.

\textit{The critical path runs through TMA, WGMMA, and barriers.} Pressure
effects inside a WarpGroup, namely bank conflicts, register spilling, and
divergence, are not modeled, and non-MMA work is charged as a per-tile
constant, so the 988-cycle budget of Section~\ref{fa3_e2e_pipe} must be
re-derived for a different precision or tile shape. The 330-cycle
per-iteration residual of Section~\ref{generic_frontend} quantifies the
remainder: Sim-FA charges no TMAU-internal issue pipeline, and the $N/2$
WGMMA law, fitted for large shapes, is optimistic inside a tight K-loop. We
report it as a slope rather than absorbing it into the fixed intercept.

\textit{The memory hierarchy is the calibrated resource.} The remote-copy
proxy, XOR slice hash, and MSHR count are behavioral surrogates for
undisclosed Hopper mechanisms, calibrated once against H800; a study that
redesigns L2 partitioning or request tracking must re-derive them. Only dense
FP16 WGMMA is modeled, so FP8 requires a calibrated latency law rather than a
different execution model.

\textit{The schedule comes from a real execution.} Design-space exploration
is sound only while the software schedule is fixed, and SimFA-python covers
the rest at the cost of the inter-SM timing effects of
Section~\ref{analytical_exp}. Sim-FA is kernel-level: one GPU, one kernel, a
cold L2, with no interconnect or multi-tenant effects. We validate only the
non-causal configuration; the core imposes no such restriction, but we do not
claim validated accuracy there.

\section{Related Work}
Table ~\ref{tab:related-comparison} summarizes representative GPU/LLM-oriented simulators and performance models across key capability dimensions. Classical accelerator-level analytical and DSE frameworks are discussed separately below.

\begin{table}[t]
\centering
\caption{Comparison with representative GPU and LLM modeling tools.}
\vspace*{-0.8\baselineskip}
\label{tab:related-comparison}
\scriptsize
\renewcommand{\arraystretch}{1.08}
\setlength{\tabcolsep}{1.8pt}   

\begin{tabular*}{\columnwidth}
{@{\extracolsep{\fill}}lcccccc@{}}
\toprule
\textbf{Dim.} &
\textbf{A-Sim} &
\textbf{MGP} &
\textbf{GenZ/LV} &
\textbf{Compass} &
\textbf{GTSim} &
\textbf{Sim-FA} \tabularnewline
\midrule

Type
  & cyc.
  & cyc.
  & ana.
  & ana.
  & tile
  & cyc.+ana. \tabularnewline

TMA/WG
  & \checkmark
  & N/A
  & N/A
  & --
  & abstr.
  & \checkmark \tabularnewline

Warp spec.
  & \checkmark
  & --
  & --
  & --
  & \checkmark
  & \checkmark \tabularnewline

FlashAttn.
  & \checkmark
  & --
  & --
  & --
  & \checkmark
  & \checkmark \tabularnewline

Memory
  & detailed
  & cyc.
  & BW
  & BW
  & tile
  & L2+DRAM \tabularnewline

Scope
  & \shortstack{general\\GPU}
  & \shortstack{AMD\\mGPU}
  & \shortstack{LLM\\DSE}
  & \shortstack{LLM\\DSE}
  & ML
  & \shortstack{async.\\ML} \tabularnewline

\shortstack[l]{Long-seq.\\DRAM}
  & cyc.
  & cyc.
  & ideal.
  & N/A
  & implicit
  & cap.-aware \tabularnewline

\bottomrule
\end{tabular*}
\end{table}

\subsection{Cycle-accurate Simulators}
Detailed GPU simulators provide high-fidelity execution models for a broad range of GPU workloads. Accel-Sim~\cite{AccelSim,accelsim2} is a widely used and extensively validated trace-driven GPU simulation framework. Its recent 2.0 release extends the framework to NVIDIA Hopper GPUs, including TMA, asynchronous WGMMA, mbarrier synchronization, and the corresponding memory-system extensions. 
Other detailed simulators target different substrates: MGPUSIM~\cite{mgpusim}
for multi-GPU AMD systems and SCALE-Sim~\cite{scalesim} for systolic-array
accelerators.

Sim-FA therefore differs from Accel-Sim not primarily in feature availability, but in simulation abstraction and scope. Accel-Sim retains a detailed, general-purpose SASS-driven execution model intended to cover a broad range of GPU workloads. In contrast, Sim-FA adopts a workload-focused WarpGroup-level abstraction for asynchronous ML kernels. It explicitly preserves the timing-visible interactions among TMA transfers, WGMMA execution, barrier synchronization, producer--consumer scheduling, and the memory hierarchy, while omitting lower-level pipeline details that are not required to reproduce these interactions. This specialization is designed to reduce simulation time and host-memory cost for TMA/WGMMA-intensive kernels while retaining cycle-level visibility into their asynchronous pipeline behavior.

\subsection{Tile-granularity Simulators}
GPU-Tile-Sim (GTSim)~\cite{gtsim} represents kernel execution as a warp-level tile graph whose nodes are tile-level operations and whose edges encode data and ordering constraints, based on the insight that LLM kernel performance is governed less by instruction latency than by
dependency structure. This graph-driven approach captures software pipelining and compute--memory overlap at low cost, reporting 1.22\%--8.71\% MAPE on GEMM and attention. However, since it omits cycle-accurate memory modeling, GTSim cannot be used to verify architectural innovations in memory hierarchies.


\subsection{Analytical Models}
Analytical models are widely used for design-space exploration (DSE) as they provide much higher exploration speed than detailed simulation. Timeloop~\cite{timeloop} systematically evaluates DNN accelerator architectures and mappings by modeling loop transformations, data movement, and hierarchical storage. MAESTRO~\cite{maestro_micro2019} adopts a data-centric representation to analytically estimate reuse, communication, performance, and hardware cost under different dataflows. 

More recent analytical models directly target LLM inference. GenZ~\cite{genz} explores LLM workloads across hardware and system configurations, while LLM Viewer~\cite{llm_viewer} uses roofline-style analysis to estimate computation and memory requirements. LLM Compass~\cite{compass} further combines performance modeling with architecture exploration. Such high-level abstractions are well suited to large DSE studies, but do not explicitly capture fine-grained asynchronous execution or all timing-sensitive memory-hierarchy effects.

As discussed in Section ~\ref{analytical_exp}, this abstraction can become inaccurate for long-sequence attention: GenZ and LLM Viewer may underestimate DRAM traffic when reusable K/V data exceed the effective LLC capacity, while LLM Compass does not model FlashAttention. SimFA-python instead uses a concurrency-aware traffic formulation and shares the same memory-traffic model with the cycle-level Sim-FA framework. We therefore use analytical modeling as a fast path for large-scale DSE, while retaining cycle-level simulation when asynchronous dependencies and memory-system interactions become first-order performance effects.

\section{Conclusion}
Modern GPU kernels increasingly rely on WarpGroup specialization, asynchronous TMA transfers, WGMMA execution, and fine-grained synchronization. Sim-FA occupies a complementary design point to general-purpose detailed GPU simulators: it preserves the timing-visible interactions required by asynchronous ML pipelines while omitting lower-level state that is not necessary for the target analysis. Sim-FA reproduces FlashAttention-3 latency on H800 with 5.7\% MAPE, while the operator-agnostic TTGIR frontend achieves 5.49\% MAPE across 23 GEMM evaluation shapes using the same core and memory-system configuration. On operator-equivalent GEMM workloads, Sim-FA is faster than Accel-Sim 2.0 in 22 of 23 cases, with a \(7.4\times\) geometric-mean simulation speedup. 

Finally, SimFA-python complements the cycle-level core with a fast analytical path and captures capacity-induced K/V refetching that becomes important for long-sequence attention.

\bibliographystyle{IEEEtranS}
\bibliography{refs}

\newpage

\section{Biography Section}
\begin{IEEEbiography}[{\includegraphics[width=1in,height=1.25in,clip,keepaspectratio]{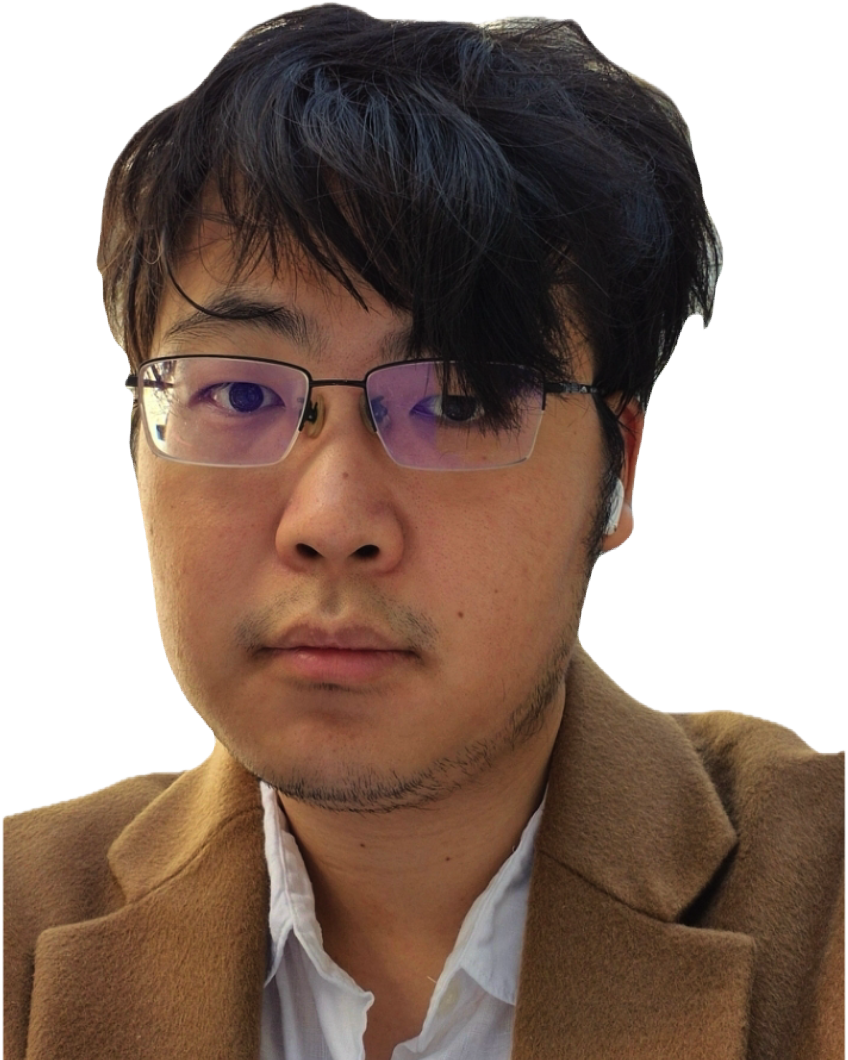}}]{Zhongchun Zhou}
 received the B.E. degree in microelectronic engineering from the School of Integrated Circuits, Tsinghua University, Beijing, China, in 2022. He is currently pursuing the Ph.D. degree in electronic and computer engineering with the Hong Kong University of Science and Technology, Hong Kong. His research interests include integrated circuits, field-programmable gate arrays (FPGAs), high performance computing (HPC) and computer architecture, with a focus on hardware acceleration and system optimization for large language models.
\end{IEEEbiography}
\begin{IEEEbiography}[{\includegraphics[width=1in,height=1.25in,clip,keepaspectratio]{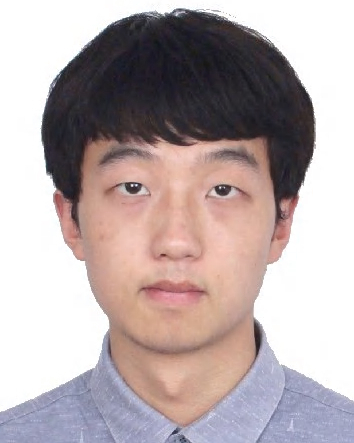}}]{Yuhang Gu}
is currently an undergraduate student from the School of Electronic Science and Engineering, Southeast University, Nanjing, China, since 2022. He is interested in research on computer systems, advanced computing systems including high performance computing and quantum computing, integrated circuits and computer architecture.
\end{IEEEbiography}
\begin{IEEEbiography}[{\includegraphics[width=1in,height=1.25in,clip,keepaspectratio]{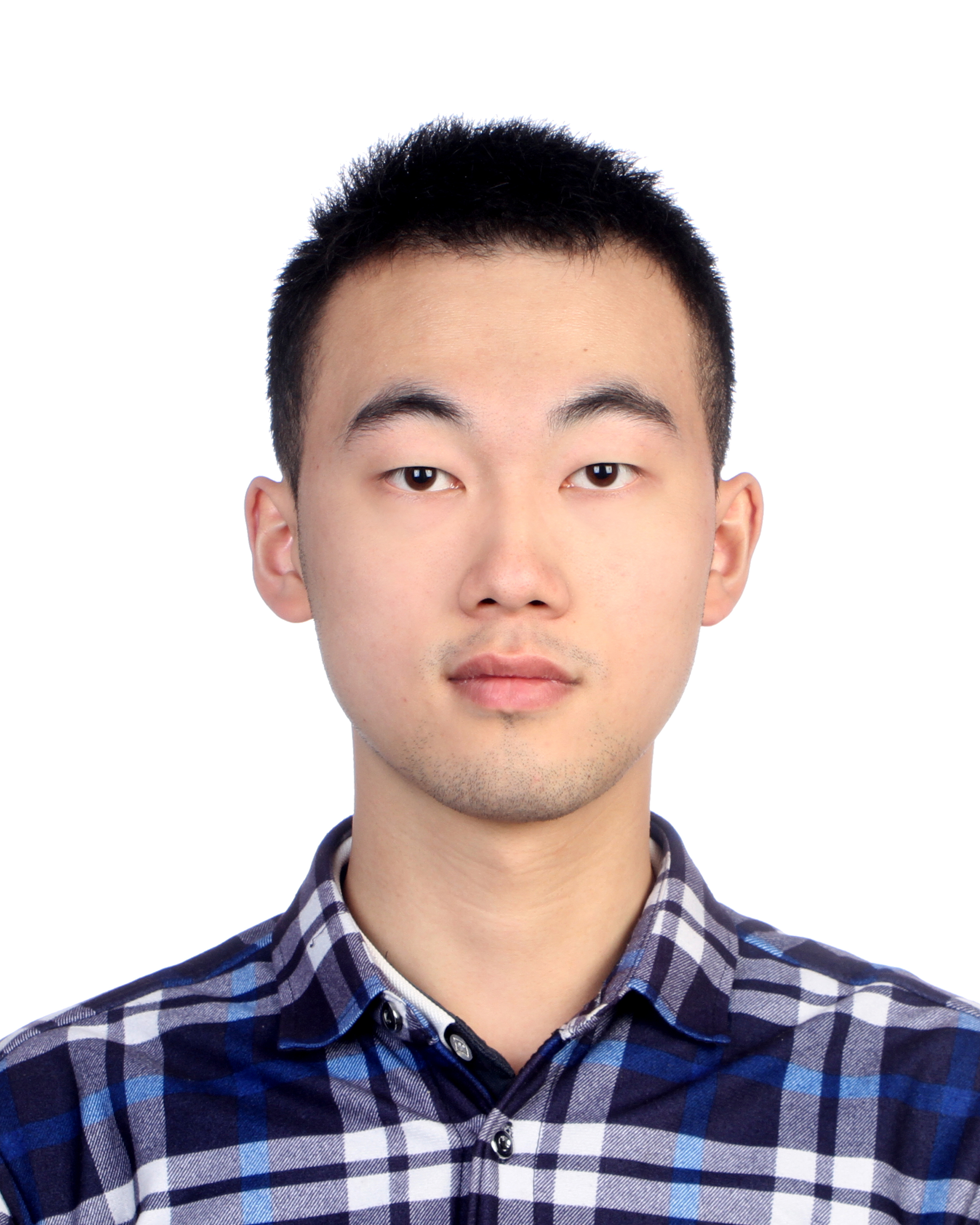}}]{Chengtao Lai}
 received the B.E. degree in Engineering Mechanics from the School of Aerospace Engineering, Tsinghua University, Beijing, China, in 2021. He is currently pursuing the Ph.D. degree in Electronic and Computer Engineering with the Hong Kong University of Science and Technology, Hong Kong. His research interests include high performance computing and computer architecture, especially the memory subsystem of AI accelerators.
\end{IEEEbiography}

\begin{IEEEbiography}[{\includegraphics[width=1in,height=1.25in,clip,keepaspectratio]{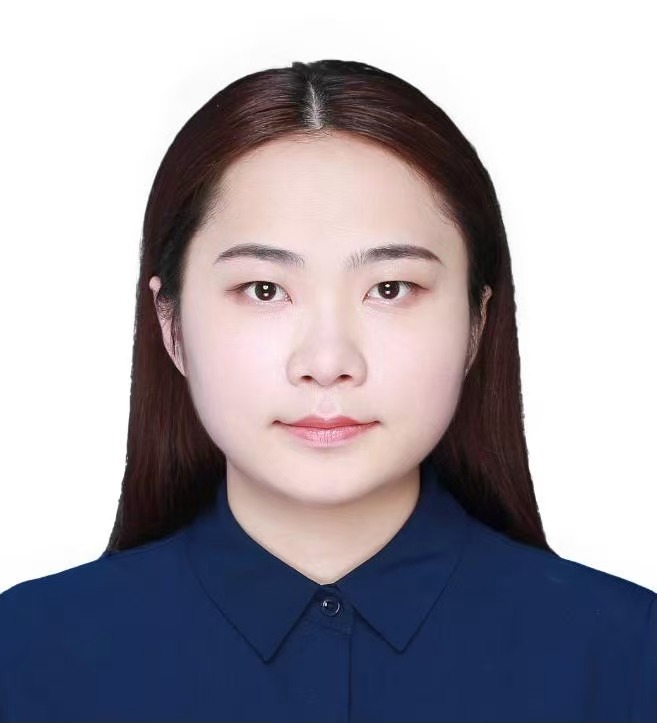}}]{Ya Wang}
received the B.S. degree in electronic information science and technology from Nanjing University, Nanjing, China, in 2022. She is currently pursuing the Ph.D. degree in electronic and computer engineering with the Hong Kong University of Science and Technology, Hong Kong. Her research interests include CPU and NPU architecture modeling and design space exploration (DSE).
\end{IEEEbiography}
\begin{IEEEbiography}[{\includegraphics[width=1in,height=1.25in,clip,keepaspectratio]{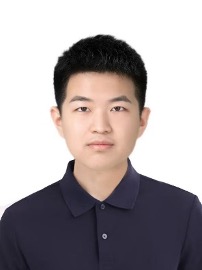}}]{Zeyu Han}
 is currently an undergraduate student from the School of Integrated Circuits, Nanjing University, Nanjing, China, since 2023. He is interested in research on computer architecture, software–hardware co-design, LLM inference systems and AI infrastructure.
\end{IEEEbiography}

\begin{IEEEbiography}[{\includegraphics[width=1in,height=1.25in,clip,keepaspectratio]{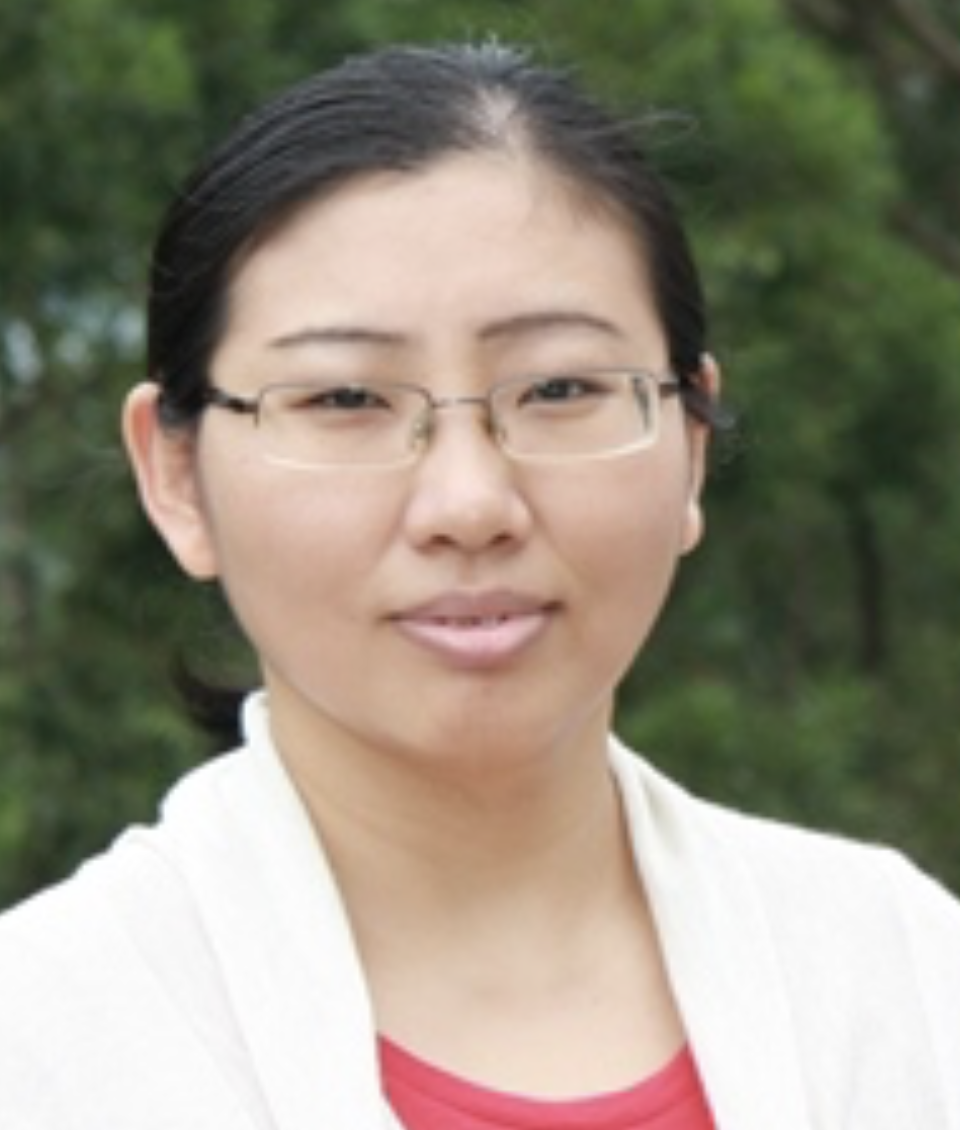}}]{Wei Zhang}
(Fellow, IEEE) received the Ph.D. degree from Princeton University, Princeton, NJ, USA, in 2009. She was an Assistant Professor with the School of Computer Engineering, Nanyang Technological University, Singapore, from 2010 to 2013. She joined the Hong Kong University of Science and Technology, Hong Kong, in 2013, where she is currently a Professor and she established the Reconfigurable Computing System Laboratory. She has authored or coauthored over 80 book chapters and papers in peer-reviewed journals and international conferences. Her current research interests include reconfigurable systems, FPGA-based design, low-power high-performance multicore systems, electronic design automation, embedded systems, and emerging technologies. Dr. Zhang serves as an Associate Editor for ACM Transactions on Embedded Computing Systems, IEEE Transactions on Very Large Scale Integration (VLSI) Systems, and ACM Journal of Emerging Technologies in Computing Systems. She also serves on many organization committees and technical program committees.
\end{IEEEbiography}

\begin{IEEEbiography}[{\includegraphics[width=1in,height=1.25in,clip,keepaspectratio]{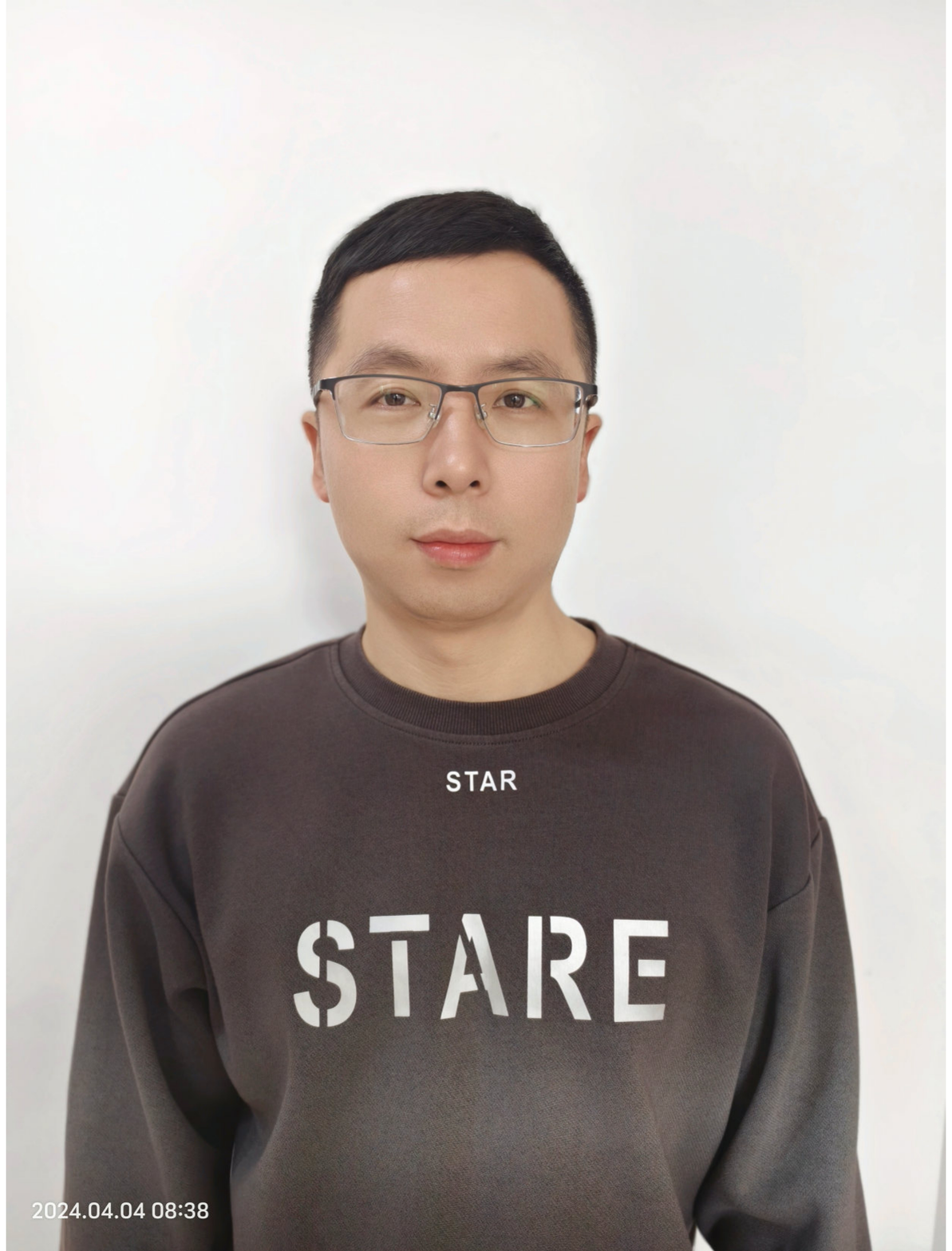}}]{Jun Liu}
is a Senior Wireless Expert at ZTE Corporation; Member of the State Key Laboratory of Mobile Network and Mobile Multimedia Technology. His current research interests include deep learning algorithms, AI compilers, AI accelerator architecture design, and simulation.
\end{IEEEbiography}

\vfill

\end{document}